\documentclass[10pt,twocolumn,twoside]{IEEEtran}
\usepackage[noadjust]{cite}

\ifCLASSINFOpdf
  \usepackage[pdftex]{graphicx}
\else
  \usepackage[dvips]{graphicx}
\fi
\usepackage[cmex10]{amsmath}
\usepackage{amssymb}
\usepackage{algorithmic,algorithm}

\usepackage[caption=false,font=footnotesize]{subfig}
\usepackage{fixltx2e}

\usepackage{multirow}

\usepackage{url}

\usepackage{dsfont}

\def\I{\ensuremath{\mathds{1}}}

\def\R{\ensuremath{\mathbb{R}}}

\def\X{\ensuremath{\mathcal{X}}}
\def\V{\ensuremath{\mathcal{V}}}
\def\Y{\ensuremath{\mathcal{Y}}}

\def\trans{\ensuremath{^\mathsf{T}}}

\def\th{\ensuremath{^\text{th}}}

\DeclareMathOperator{\trace}{tr}

\DeclareMathOperator*{\argmin}{argmin}
\DeclareMathOperator{\suchthat}{s.t.}

\def\eg{\emph{e.g.\/}}
\def\ie{\emph{i.e.\/}}
\def\etc{\emph{etc\/}}
\def\etal{\emph{et al\/}}

\usepackage{color}
\definecolor{dkgreen}{rgb}{0.0,0.6,0.0}
\definecolor{purple}{rgb}{0.6,0.0,0.8}

\begin{document}
%
\title{Learning content similarity for music recommendation}
%
%
%

\author{Brian~McFee,~\IEEEmembership{Student member,~IEEE,}
        Luke~Barrington,~\IEEEmembership{Student member,~IEEE,}
        and~Gert~Lanckriet,~\IEEEmembership{Member,~IEEE}
\thanks{B. McFee is with the Department
of Computer Science and Engineering, University of California at San Diego, La Jolla,
CA, 92093 USA (e-mail: bmcfee@cs.ucsd.edu).}
\thanks{L. Barrington and G. Lanckriet are with the Department of Electrical and Computer Engineering, University of California at San Diego, La Jolla, CA,
92093 USA (e-mail lukeinusa@gmail.com; gert@ece.ucsd.edu).}
}

%
%

\markboth{Journal of \LaTeX\ Class Files,~Vol.~6, No.~1, January~2007}%
{Shell \MakeLowercase{\textit{et al.}}: Bare Demo of IEEEtran.cls for Journals}
%



\maketitle

\begin{abstract}
Many tasks in music information retrieval, such as recommendation, and playlist generation for online radio, fall naturally into the \emph{query-by-example}
setting, wherein a user queries the system by providing a song, and the system responds with a list of relevant or similar song recommendations.  Such
applications ultimately depend on the notion of \emph{similarity} between items to produce high-quality results.  Current state-of-the-art systems employ
\emph{collaborative filter} methods to represent musical items, effectively comparing items in terms of their constituent users.  While collaborative filter
techniques perform well when historical data is available for each item, their reliance on historical data impedes performance on novel or unpopular items.
To combat this problem, practitioners rely on content-based similarity, which naturally extends to novel items, but is typically out-performed by 
collaborative filter methods.

In this article, we propose a method for optimizing content-based similarity by learning from a sample of collaborative filter data.  The optimized 
content-based similarity metric can then be applied to answer queries on novel and unpopular items, while still maintaining high recommendation accuracy.  The
proposed system yields accurate and efficient representations of audio content, and experimental results show significant improvements in accuracy over competing
content-based recommendation techniques.
\end{abstract}


\begin{IEEEkeywords}
Audio retrieval and recommendation, music information retrieval, query-by-example, collaborative filters, structured prediction.
\end{IEEEkeywords}

\begin{center} \bfseries EDICS Category: AUD-CONT \end{center}
%
\IEEEpeerreviewmaketitle

\section{Introduction}



\IEEEPARstart{A}{n} effective notion of similarity forms the basis of many applications involving multimedia data.  For example, an online music 
store can benefit greatly from the development of an accurate method for automatically assessing similarity between two songs, which can in turn 
facilitate high-quality recommendations to a user by finding songs which are similar to her previous purchases or preferences.  More generally, 
high-quality similarity can benefit any \emph{query-by-example} recommendation system, wherein a user presents an example of an item that she likes, and the
system responds with, \eg, a ranked list of recommendations.

The most successful approaches to a wide variety of recommendation tasks --- including not just music, but books, movies, \etc. --- 
is \emph{collaborative filters} (CF).
Systems based on collaborative filters exploit the ``wisdom of crowds'' to infer similarities between items, and recommend new items to users by representing and comparing these items in terms of the people who use them~\cite{goldberg1992}.
Within the domain of music information retrieval, recent studies have shown that CF systems consistently outperform alternative methods for 
playlist generation~\cite{genius} and semantic annotation~\cite{kim09}.  
However, collaborative filters suffer from the dreaded ``cold start'' problem: a new item cannot be recommended until it has been purchased, 
and it is less likely to be purchased if it is never recommended.  Thus, only a tiny fraction of songs may be recommended, making it difficult 
for users to explore and discover new music~\cite{celma2010}.

The cold-start problem has motivated researchers to improve \emph{content-based} recommendation engines.  
Content-based systems operate on music representations that are extracted automatically from the audio content,
eliminating the need for human feedback and annotation when computing
similarity.
While this approach naturally extends to any item regardless of popularity, the construction of features and definition of \emph{similarity} in these systems are 
frequently ad-hoc and not explicitly optimized for the specific task.  

\begin{figure}
\centering\includegraphics[width=0.475\textwidth]{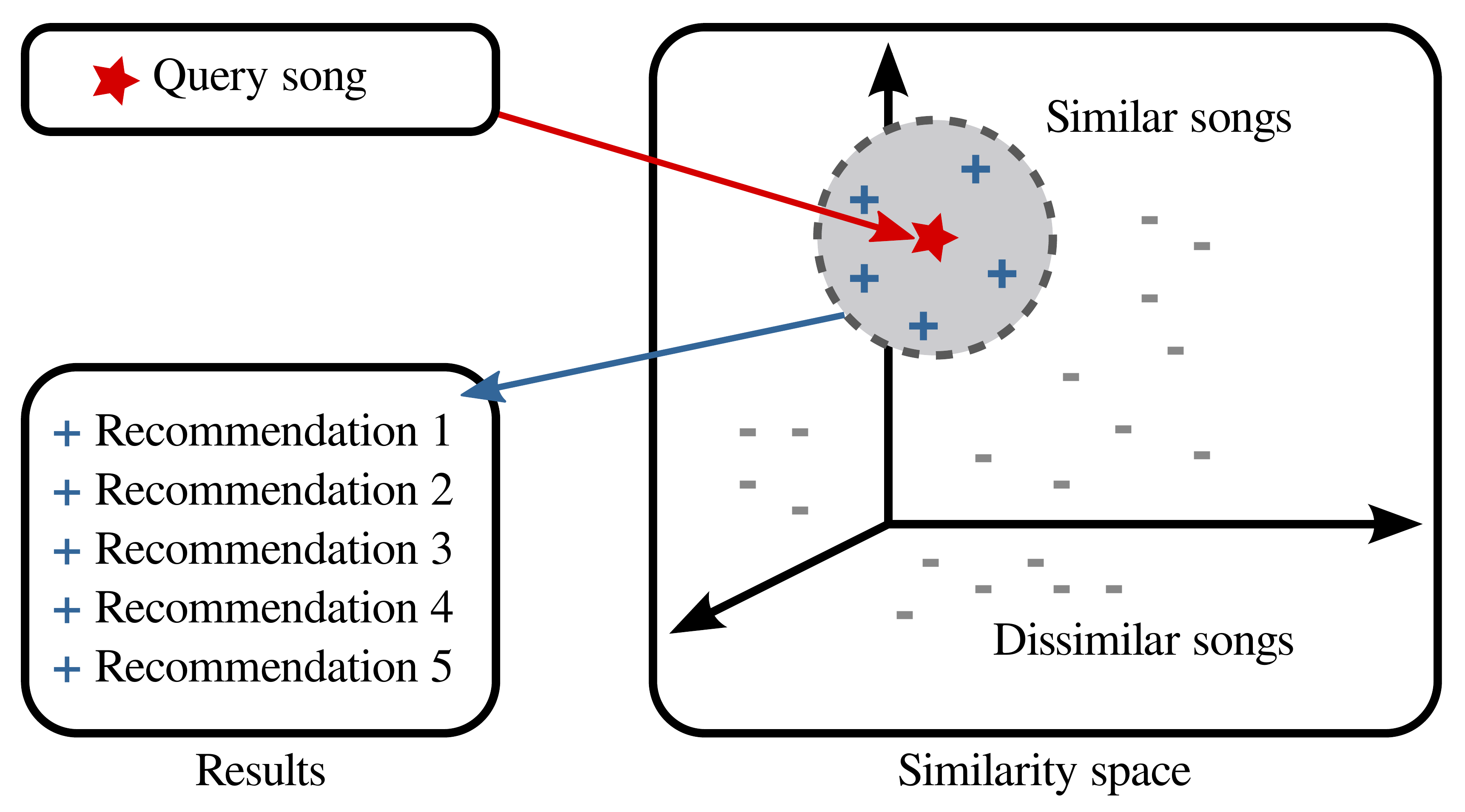}%
\caption{Query-by-example recommendation engines allow a user to search for new items by providing an example item.  Recommendations are formed by computing
the most similar items to the query item from a database of potential recommendations.\label{fig:querybyexample}}
\end{figure}

In this paper, we propose a method for optimizing content-based audio similarity by learning from a sample of collaborative filter data. Based on this optimized similarity measure, recommendations can then be made where no collaborative filter data is available. The proposed method treats similarity learning as an information retrieval problem, where similarity is learned to optimize
the ranked list
of results in response to a query example (Figure~\ref{fig:querybyexample}).  Optimizing similarity for ranking 
requires more sophisticated machinery than, \eg, genre classification for semantic search.  However, the information retrieval approach offers 
a few key advantages, which we believe are crucial for realistic music applications.  First, there are no assumptions of transitivity or 
symmetry in the proposed method.  This allows, for example, that ``The Beatles'' may be considered a relevant result for ``Oasis'', but not vice versa.
Second, CF data can be collected \emph{passively} from users by mining their listening histories, thereby directly capturing their listening habits.  
Finally, optimizing similarity for ranking directly attacks the main quantity of interest: the ordered list of retrieved items, rather 
than coarse abstractions of similarity, such as genre agreement.

\subsection{Related work}
Early studies of musical similarity followed the general strategy of first devising a model of audio content (\eg, spectral clusters~\cite{logan2001} or 
Gaussian mixture models~\cite{aucouturier02}), applying some reasonable distance function (\eg, earth-mover's distance or Kullback-Leibler divergence), and then
evaluating the proposed similarity model against some source of ground truth.  Logan and Salomon~\cite{logan2001} and Aucouturier and 
Pachet~\cite{aucouturier02} evaluated against three notions of similarity between songs: same artist, same genre, and human survey data.  
Artist or genre agreement entail strongly binary notions of similarity, which due to symmetry and transitivity may be unrealistically coarse in 
practice.  Survey data can encode subtle relationships between items, for example, triplets of the form \emph{``A is more similar to B than to
C''}~\cite{aucouturier02,ellis02,berenzweig2004}.  However, the expressive power of human survey data comes at a cost: while artist or genre 
meta-data is relatively inexpensive to collect for a set of songs, similarity survey data may require human feedback on a quadratic (for pairwise ratings) 
or cubic (for triplets) number of comparisons between songs.

Later work in musical similarity approaches the problem in the context of supervised learning: given a set of training items (songs), and some knowledge of
similarity across those items, the goal is to \emph{learn} a similarity (distance) function that can predict pairwise similarity.  Slaney~\etal.~\cite{slaney08}
derive similarity from web-page co-occurrence, and evaluate several supervised and unsupervised algorithms for learning distance metrics.  McFee and
Lanckriet~\cite{mcfee2011} develop a metric learning algorithm for triplet comparisons as described above.  Our proposed method follows in this line of work, but
is designed to optimize structured ranking loss (not just binary or triplet predictions), and uses a collaborative filter as the source of ground truth.

The idea to learn similarity from a collaborative filter follows from a series of positive results in music applications.  Slaney and White~\cite{slaney07}
demonstrate that an item-similarity metric derived from rating data matches human perception of similarity better than a content-based method.  Similarly, it has
been demonstrated that when combined with metric learning, collaborative filter similarity can be as effective as semantic tags for predicting survey
data~\cite{mcfee2011}.  Kim~\etal.~\cite{kim09} demonstrated that collaborative filter similarity vastly out-performs content-based methods for predicting
semantic tags.  Barrington~\etal.~\cite{genius} conducted a user survey, and concluded that the iTunes Genius playlist algorithm (which is at least partially
based on collaborative filters\footnote{\url{http://www.apple.com/pr/library/2008/09/09itunes.html}}) produces playlists of equal or higher quality
than competing 
methods based on acoustic content or meta-data.

Finally, there has been some previous work addressing the cold-start problem of collaborative filters for music recommendation by integrating audio 
content.  Yoshii~\etal.~\cite{yoshii08} formulate a joint probabilistic model of both audio content and collaborative filter data in order to predict user
ratings of songs (using either or both representations), 
whereas our goal here is to use audio data to predict the similarities derived from a collaborative filter.  Our problem setting is most
similar to that of Stenzel and Kamps~\cite{stenzel2005}, wherein a CF matrix was derived from playlist data, clustered into latent ``pseudo-genres,'' and 
classifiers were trained to predict the cluster membership of songs from audio data.  Our proposed setting differs in that we derive 
similarity at the user level (not playlist level), and automatically learn the content-based song similarity that directly optimizes the primary quantity of interest in an information retrieval system: the quality of the rankings it induces.

\subsection{Our contributions}
Our primary contribution in this work is a framework for improving content-based audio similarity by learning from a sample of collaborative filter data.  Toward
this end, we first develop a method for deriving item similarity from a sample of collaborative filter data.
We then use the sample similarity to train an optimal distance metric over audio descriptors. More precisely, a distance metric is optimized to produce high-quality rankings of the training sample in a query-by-example setting.
The resulting distance metric can then be applied to previously unseen data for which collaborative filter data
is unavailable.  Experimental results verify that the proposed methods significantly outperform competing methods for content-based music retrieval.

\subsection{Preliminaries}
For a $d$-dimensional vector $u\in\R^d$ let $u[i]$ denote its $i\th$ coordinate; similarly, for a matrix $A$, let $A[ij]$ denote its $i\th$ row and $j\th$ column
entry.  A square, symmetric matrix ${A\in\R^{d\times d}}$ is \emph{positive semi-definite} (PSD, denoted $A\succeq0$) if each of its eigenvalues is 
non-negative.  For two matrices $A,B$ of compatible dimension, the Frobenius inner product is defined as
\[
\langle A, B\rangle_\text{F} = \trace(A\trans B) = \sum_{i,j} A[ij]B[ij].
\]
Finally, let $\I[x]$ denote the binary indicator function of the event $x$.

\section{Learning similarity}

The main focus of this work is the following information retrieval problem: given a 
\emph{query} song $q$, 
return a ranked list from a database $\X$ of $n$
songs ordered by descending similarity to $q$.  In general, the query may be previously unseen to the system,
but $\X$ will remain fixed across all queries. 
We will assume that each song is represented by a vector in $\R^d$, and similarity is computed by Euclidean 
distance.  Thus, for any query $q$, a natural ordering of $x\in\X$ is generated by sorting according to increasing distance from $q$: $\|q-x\|$. 

Given some side information describing the similarity relationships between items of $\X$, distance-based ranking can be improved 
by applying a \emph{metric learning} algorithm.  Rather than rely on native Euclidean distance, the learning algorithm produces a PSD
matrix $W\in\R^{d\times d}$ which characterizes an optimized distance:
\begin{equation}
\|q-x\|_W = \sqrt{(q-x)\trans W (q-x)}\label{eq:mahalanobis}.
\end{equation}
In order to learn $W$, we will apply the metric learning to rank (MLR)~\cite{mcfee10a} algorithm (Section~\ref{sec:learning:mlr}).  At a high level, MLR
optimizes the distance metric $W$ on $\X$, \ie, so that $W$ generates optimal rankings of songs in $\X$ when using each song in $\X$ as a query.
To apply the algorithm, we must provide a set of similar
songs $x \in \X$ for each \emph{training} query $q \in \X$.
This is achieved by leveraging the side information that is available for items in $\X$. More specifically, we will derive a notion of similarity from collaborative filter data on $\X$. So, the proposed approach optimizes content-based audio similarity by learning from a sample of collaborative filter data.

\subsection{Collaborative filters}
\label{sec:cfsim}

The term \emph{collaborative filter} (CF) is generally used to denote to a wide variety of techniques for modeling the interactions between a set of items 
and a set of users~\cite{goldberg1992,sarwar2001}.
Often, these interactions are modeled as a (typically sparse) matrix $F$ where rows represent the users, and columns represent the items.  The entry $F[ij]$
encodes the interaction between user $i$ and item $j$. 

The majority of work in the CF literature deals with $F$ derived from explicit user feedback, \eg, 5-star ratings~\cite{slaney07,yoshii08}.  While rating data
can provide highly accurate representations of user-item affinity, it also has drawbacks, especially in the domain of music.  First, explicit ratings require 
active participation on behalf of users.  This may be acceptable for long-form content such as films, in which the time required for a user to rate an item 
is miniscule relative to the time required to consume it.  However, for short-form content (\eg, songs), it seems unrealistic to expect a user to rate even a
fraction of the items consumed.  Second, the scale of rating data is often arbitrary, skewed toward the extremes (\eg, 1- and 5-star ratings), 
and may require careful 
calibration to use effectively~\cite{slaney07}.

Alternatively, CF data can also be derived from \emph{implicit} feedback.  While somewhat noisier on a per-user basis than explicit feedback, implicit 
feedback can be derived in much higher volumes by simply counting how often a user interacts with an item (\eg, 
listens to an artist)~\cite{deshpande2004,hu2008}.  Implicit feedback differs from rating data, in that it is positive and unbounded, and it 
does not facilitate explicit negative feedback.  As suggested by Hu~\etal.~\cite{hu2008}, binarizing an implicit feedback 
matrix by thresholding can provide an effective mechanism to infer positive associations.  

In a binary CF matrix $F$, each column $F[\cdot j]$ can be interpreted as a \emph{bag-of-users} representation of item $j$.  Of central interest in this paper 
is the similarity between items (\ie, columns of $F$).  We define the similarity between two items $i, j$ as the Jaccard index~\cite{jaccard1901} of 
their user sets:
\begin{equation}
S(i,j) = \frac{|F[\cdot i]\cap F[\cdot j]|}{|F[\cdot i] \cup F[\cdot j]|} = \frac{F[\cdot i]\trans F[\cdot j]}{|F[\cdot i]| + |F[\cdot j]| - F[\cdot i]\trans F[\cdot j]},\label{eq:jaccard}
\end{equation}
which counts the number of users shared between $A$ and $B$, and normalizes by the total number of users for $A$ or $B$.

Equation~\eqref{eq:jaccard} defines a quantitative metric of similarity between two items.  However, for information retrieval applications, we are primarily
interested in the most similar (relevant) items for any query.  We therefore define the \emph{relevant} set $\X_q^+$ for any item $q$ as the top $k$ most 
similar items according to Equation~\eqref{eq:jaccard}, \ie, those items which a user of the system would be shown first.  Although binarizing similarity in this
way does simplify the notion of relevance, it still provides a flexible language for encoding relationships between items.  Note that after thresholding, 
transitivity and symmetry are not enforced, so it is possible, \eg, for \emph{The Beatles} to be relevant for \emph{Oasis} but not vice versa.  Consequently, we 
will need a learning algorithm which can support such flexible encodings of relevance.

\subsection{Metric learning to rank}
\label{sec:learning:mlr}

Any query-by-example retrieval system must have at its core a mechanism for comparing the query to a known database, \ie, assessing similarity (or distance).
Intuitively, the overall system should yield better results if the underlying similarity mechanism is optimized according to the chosen task.  In classification
tasks, for example, this general idea has led to a family of algorithms collectively known as \emph{metric learning}, in which a feature space is optimized 
(typically by a linear transformation) to improve performance of nearest-neighbor classification~\cite{xing2003,weinberger2006,davis2007}.  While metric 
learning algorithms have
been demonstrated to yield substantial improvements in classification performance, nearly all of them are fundamentally limited to classification, and do not
readily generalize to asymmetric and non-transitive notions of similarity or relevance.  Moreover, the objective functions optimized by most metric learning
algorithms do not clearly relate to ranking performance, which is of fundamental interest in information retrieval applications.

Rankings, being inherently combinatorial objects, can be notoriously difficult to optimize.  Performance measures of rankings, \eg, area under the ROC curve
(AUC)~\cite{egan1975}, are typically non-differentiable, discontinuous functions of the underlying parameters, so standard numerical optimization 
techniques cannot be directly applied.  However, in recent years, algorithms based on the structural SVM~\cite{tsochantaridis06} have been developed which can
efficiently optimize a variety of ranking performance measures~\cite{joachims2005,yue2007,chakrabarti2008}.  While these algorithms support general notions 
of relevance, they do not directly exploit the structure of query-by-example retrieval problems. 

The metric learning to rank (MLR) algorithm combines these two approaches of metric learning and structural SVM, and is designed specifically for the
query-by-example setting~\cite{mcfee10a}.  
MLR learns a positive semi-definite matrix $W$ such 
that rankings induced by learned distances (Equation~\eqref{eq:mahalanobis}) are optimized according to a ranking loss measure, \eg, AUC, mean 
reciprocal rank (MRR)~\cite{voorhees2001}, or normalized discounted cumulative gain (NDCG)~\cite{jarvelin2000}.  
In this setting, ``relevant'' results should lie close in space to the query $q$, and ``irrelevant'' results should 
be pushed far away.

\begin{figure}
\centering%
\includegraphics[width=0.45\textwidth]{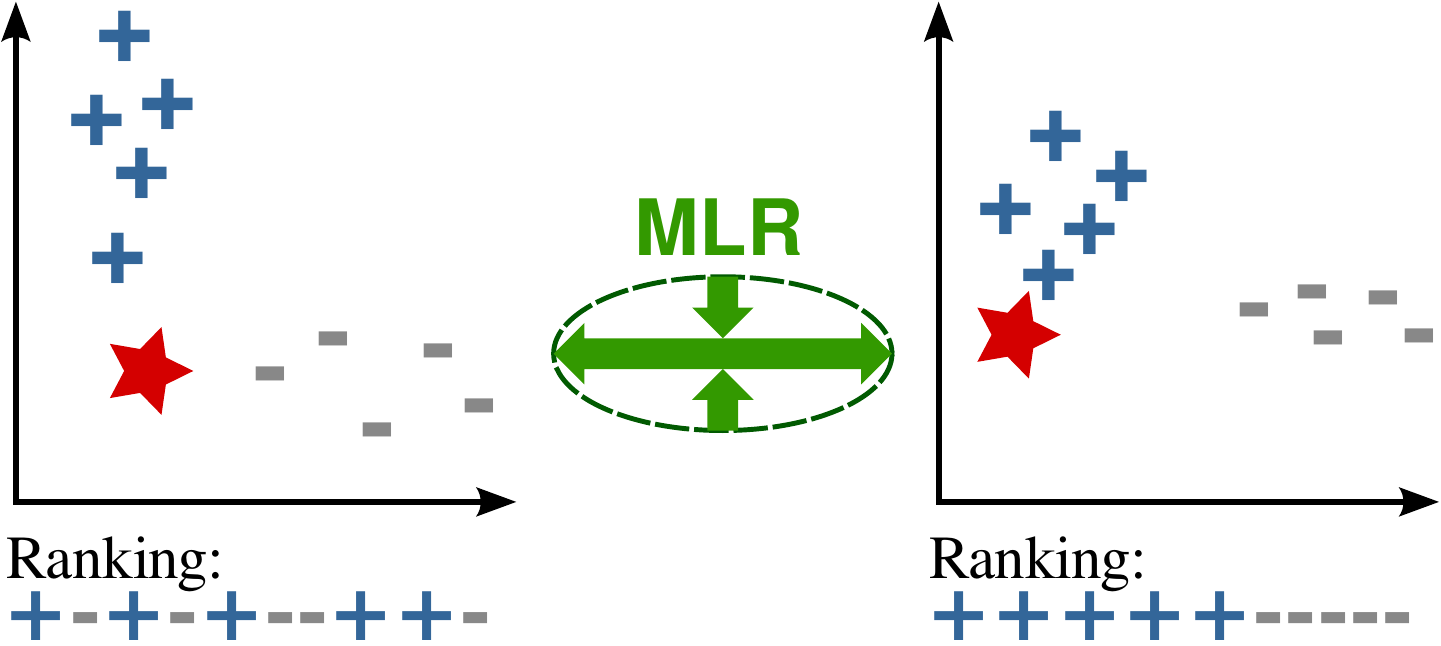}
\caption{Left: a query point $\star$ and its relevant (+) and irrelevant (-) results; ranking by distance from $\star$ results in poor retrieval performance.
Right: after learning an optimal distance metric with MLR, relevant results are ranked higher than irrelevant results.\label{fig:mlr}}
\end{figure}

For a query song $q$, the database $\X$ is ordered by sorting $x\in\X$ according to
increasing distance from $q$ under the metric defined by $W$ (see Figure~\ref{fig:mlr}).
The metric $W$ is learned by solving a constrained convex optimization problem such that, for each training query $q$, a higher score is assigned to a correct ranking $y_q$ than to any other ranking $y\in\Y$ (the set of all rankings):
\begin{equation}
\forall q:~ \langle W, \psi(q, y_q) \rangle_\text{F} \geq \langle W, \psi(q,y)\rangle_\text{F} + \Delta(y_q, y) - \xi_q.\label{eq:marginconstraint}
\end{equation}
Here, the ``score'' for a query-ranking pair $(q,y)$ is computed by the Frobenius inner product $\langle W, \psi(q,y)\rangle_\text{F}$.
$\psi(q,y)$ is a matrix-valued feature map which encodes the query-ranking pair $(q,y)$, and $\Delta(y_q,y)$ computes the loss (\eg, decrease in AUC)
incurred by predicting $y$ instead of $y_q$ for the query $q$, essentially playing the role of the ``margin'' 
between rankings $y_q$ and $y$.  Intuitively, the score for a correct ranking $y_q$ should exceed the
score for any other $y$ by at least the loss $\Delta(y_q,y)$.  In the present context, a correct
ranking is any one which places all relevant results $\X_q^+$ before all irrelevant results $\X_q^-$.
To allow violations of margins during training, a slack variable $\xi_q\geq0$ is introduced for each query.

Having defined the margin constraints (Equation~\eqref{eq:marginconstraint}), what remains to be specified, to learn $W$, is the feature map $\psi$ and the objective function
of the optimization.  To define the feature map $\psi$, we first observe that the margin constraints indicate that, for a query $q$, the predicted ranking 
$y$ should be that which maximizes the score $\langle W, \psi(q,y)\rangle_\text{F}$.  Consequently, the (matrix-valued) feature map $\psi(q,y)$ must be 
chosen so that the score maximization coincides with the distance-ranking induced by $W$, which is, after all, the prediction rule we propose to use in practice, 
for query-by-example recommendation (Equation~\eqref{eq:mahalanobis}).
To accomplish this, MLR encodes query-ranking pairs $(q,y)$ by the \emph{partial order} feature~\cite{joachims2005}:
\begin{equation}
\psi(q,y) = \sum_{i\in\X_q^+} \sum_{j\in\X_q^-} y_{ij} \frac{ \left(\phi(q,i) - \phi(q,j)\right)}{|\X_q^+|\cdot|\X_q^-|},
\end{equation}
where $\X_q^+$ ($\X_q^-$) is the set of relevant (irrelevant) songs for $q$, the ranking $y$ is encoded by
\[
    y_{ij}  =   \begin{cases}
                    +1 & i \text{ before } j \text{ in } y\\
                    -1 & i \text{ after } j
                \end{cases},
\]
and $\phi(q,i)$ is an auxiliary (matrix-valued) feature map that encodes the relationship between the query $q$ and an individual result $i$.
Intuitively, $\psi(q,y)$ decomposes the ranking $y$ into pairs $(i,j) \in \X_q^+\times\X_q^-$, and computes a signed average of pairwise differences $\phi(q,i)-\phi(q,j)$.
If $y$ places $i$ before $j$ (\ie, correctly orders $i$ and $j$), the difference $\phi(q,i)-\phi(q,j)$ is added to $\psi(q,y)$, and otherwise it is
subtracted.  Note that under this definition of $\psi$, any two correct rankings $y_q, y'_q$ have the same feature representation: $\psi(q,y_q) =\psi(q,y'_q)$.
It therefore suffices to only encode a single correct ranking $y_q$ for each query $q$ to construct margin constraints (Equation~\eqref{eq:marginconstraint}) during optimization.

Since $\psi$ is linear in $\phi$, the score also decomposes into a signed average across pairs:
\begin{equation}\label{eq:score_decomposed}
\langle W, \psi(q,y)\rangle_\text{F} = \sum_{i\in\X_q^+} \sum_{j\in\X_q^-} y_{ij} \frac{ \langle W, \phi(q,i)\rangle_\text{F} - \langle W, \phi(q,j) \rangle_\text{F} }{|\X_q^+|\cdot|\X_q^-|}.
\end{equation}
This indicates that the score $\langle W, \psi(q,y_q)\rangle_\text{F}$ for a correct ranking $y_q$ (the left-hand side of Equation~\eqref{eq:marginconstraint}) will be larger when the point-wise score $\langle W, \phi(q,\cdot)\rangle_\text{F}$ is high for relevant points $i$, and low for irrelevant points $j$, \ie,
\begin{equation}\label{eq:order}
\forall i \in \X_q^+, j \in \X_q^-:~\langle W, \phi(q,i) \rangle_\text{F} > \langle W, \phi(q,j) \rangle_\text{F}.
\end{equation}
Indeed, this will accumulate only positive terms in the score computation in Equation~\eqref{eq:score_decomposed}, since a correct ranking orders all relevant results $i$ before all irrelevant results $j$ and, thus, each $y_{ij}$ in the summation will be positive.
Similarly, for incorrect rankings $y$, point-wise scores satisfying Equation~\eqref{eq:order} will lead to smaller scores $\langle W, \psi(q,y)\rangle_\text{F}$. Ideally, after training, $W$ is maximally aligned to correct rankings $y_q$ (\ie, $\langle W, \psi(q,y_q)\rangle_\text{F}$ achieves large margin over scores $\langle W, \psi(q,y)\rangle_\text{F}$ for incorrect rankings) by (approximately) satisfying Equation~\eqref{eq:order}.
Consequently, at test time (\ie, in the absence of a correct ranking $y_q$), the ranking for a query $q$ 
is predicted by sorting $i\in\X$ in descending order of point-wise score $\langle W, \phi(q,i)\rangle_\text{F}$~\cite{joachims2005}.

This motivates the choice of $\phi$ used by MLR:
\begin{equation}
\phi(q,i) = - (q-i)(q-i)\trans\label{eq:phi},
\end{equation}
which upon taking an inner product with $W$, yields the negative, squared distance between $q$ and $i$ under $W$:
\begin{align}
\langle W, \phi(q,i)\rangle_\text{F} &= -\trace\left(W(q-i)(q-i)\trans\right)\\
                            &= - (q-i)\trans W (q-i)\nonumber\\
                            &= -\|q-i\|_W^2\nonumber.
\end{align}
Descending point-wise score $\langle W, \phi(q,i)\rangle_\text{F}$ therefore corresponds to increasing distance from $q$.  As a result, the
ranking predicted by descending score is equivalent to that predicted by increasing distance from $q$, which is precisely the ranking of interest for
query-by-example recommendation.

\begin{algorithm}[!t]
\begin{algorithmic}
\REQUIRE{data $\X=\{q_1, q_2, \dots, q_n\} \subset \R^d$,\\\hspace{2.25em}correct rankings $\{y_q:~q\in\X\}$,\\ \hspace{2.25em}slack trade-off $C>0$}
\ENSURE{$d\times d$ matrix $W\succeq 0$}
\end{algorithmic}
\begin{align*}
\displaystyle\min_{W\succeq0,\xi}\hspace{1cm} & \trace(W) + C\cdot\frac{1}{n}\sum_{q\in\X} \xi_q\\
\suchthat\hspace{0.5cm}\forall q \in \X, \hspace{0.5em}&\forall y \in \Y:\\
\langle W, \psi(q, y_q)\rangle_\text{F} &\geq \langle W, \psi(q, y)\rangle_\text{F} + \Delta(y_q, y) - \xi_q
\end{align*}
\caption{Metric learning to rank~\cite{mcfee10a}\label{alg:mlr}}
\end{algorithm}

The MLR optimization problem is listed as Algorithm~\ref{alg:mlr}.  As in support vector machines~\cite{cortes1995}, 
the objective consists of two competing terms:
a regularization term $\trace(W)$, which is a convex approximation to the rank of the learned metric, and $1/n\sum\xi_q$ provides a 
convex upper bound on the empirical training loss $\Delta$, and the two terms are balanced by a trade-off parameter $C$.  Although the 
full problem includes a super-exponential number of constraints (one for each $y\in\Y$, for each $q$), it can be approximated by cutting 
plane optimization techniques~\cite{joachims09,mcfee10a}.

\section{Audio representation}

In order to compactly summarize audio signals, we represent each song as a histogram over a dictionary of timbral \emph{codewords}.  This general 
strategy has been successful in computer vision applications~\cite{fei2005bayesian}, as well as audio and music
classification~\cite{sundaram2008,seyerlehner2008,hoffman09}.
As a first step, a \emph{codebook} is constructed by clustering a large collection of feature descriptors (Section~\ref{sec:audio:codebook}).  Once the codebook
has been constructed, each song is summarized by aggregating vector quantization (VQ) representations across all frames in the song, resulting in \emph{codeword
histograms} (Section~\ref{sec:audio:softvq}).  Finally, histograms are represented in a non-linear kernel space to facilitate better learning with
MLR (Section~\ref{sec:audio:histogram}).

\subsection{Codebook training}
\label{sec:audio:codebook}
Our general approach to constructing a codebook for vector quantization is to aggregate audio feature descriptors from a large pool of songs into a single
bag-of-features, which is then clustered to produce the codebook.

For each song $x$ in the codebook training set $\X_C$ --- which may generally be distinct from the MLR training set $\X$ --- we compute the first 13 Mel
frequency cepstral coefficients (MFCCs)~\cite{rabiner1993} from each half-overlapping 23ms frame.  From the time series of MFCC vectors, we compute the first and second 
instantaneous derivatives, which are concatenated
to form a sequence of 39-dimensional dynamic MFCC ($\Delta$MFCC) vectors~\cite{buchanan2005}.  These descriptors are then aggregated
across all $x\in\X_C$ to form an unordered bag of features $Z$.

To correct for changes in scale across different $\Delta$MFCC dimensions, each vector $z\in Z$ is normalized according to the sample 
mean $\mu\in\R^{39}$ 
and standard deviation $\sigma\in\R^{39}$ 
estimated from $Z$.  The $i\th$ coordinate $z[i]$ is mapped by
\begin{equation}
z[i] \mapsto \frac{z[i] - \mu[i]}{\sigma[i]}\label{eq:zscore}.
\end{equation}
The normalized $\Delta$MFCC vectors are then clustered into a set $\V$ of $|\V|$ codewords by k-means (specifically, an online variant of Hartigan's
method~\cite{hartigan1975}).

\subsection{(Top-$\tau$) %
Vector quantization}
\label{sec:audio:softvq}

Once the codebook $\V$ has been constructed, a song $x$ is represented as a histogram $h_x$ over the codewords in $\V$.  This proceeds in three steps: 1) a
bag-of-features is computed from $x$'s $\Delta$MFCCs, denoted as $x=\{x_i\}\subset\R^{39}$; 2) each $x_i\in x$ is normalized according to 
Equation~\eqref{eq:zscore}; 3) the codeword histogram is constructed by counting the frequency with which each codeword $v\in\V$ quantizes an element of
$x$:\footnote{To simplify notation, we denote by $h_x[v]$ the bin of histogram $h_x$ corresponding to the codeword $v\in\V$.  Codewords are assumed to be unique,
and the usage should be clear from context.}
\begin{equation}
h_x[v] = \frac{1}{|x|} \sum_{x_i\in x} \I\left[v = \argmin_{u\in\V}\|x_i - u\|\right].\label{eq:vq}
\end{equation}
Codeword histograms are normalized by the number of frames $|x|$ in the song in order to ensure comparability between songs of different lengths; $h_x$ may
therefore be interpreted as a multinomial distribution over codewords.

\begin{figure}
\centering%
\includegraphics[width=0.23\textwidth]{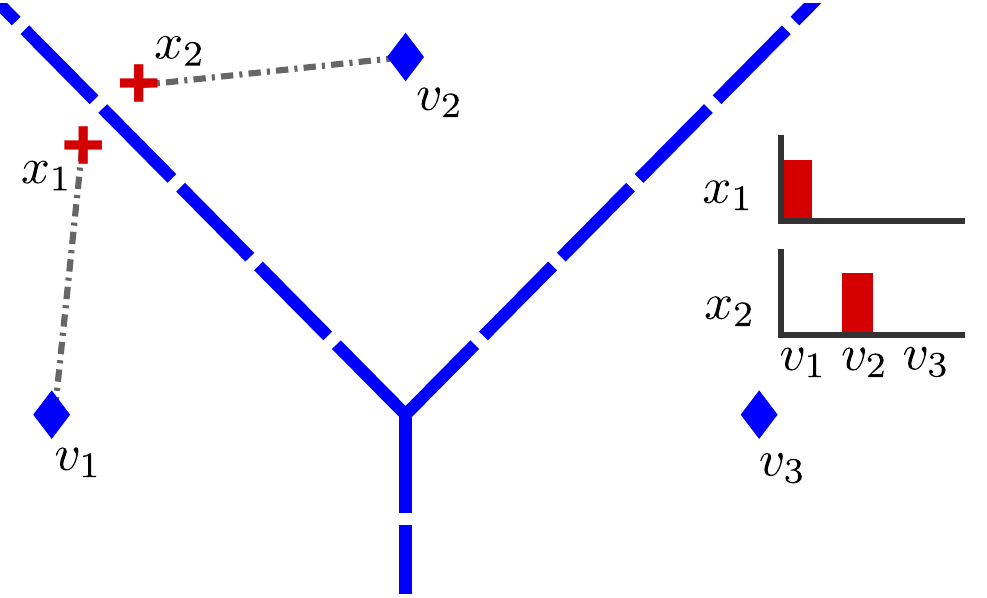}\label{fig:softvq:hard}%
\hfill\includegraphics[width=0.23\textwidth]{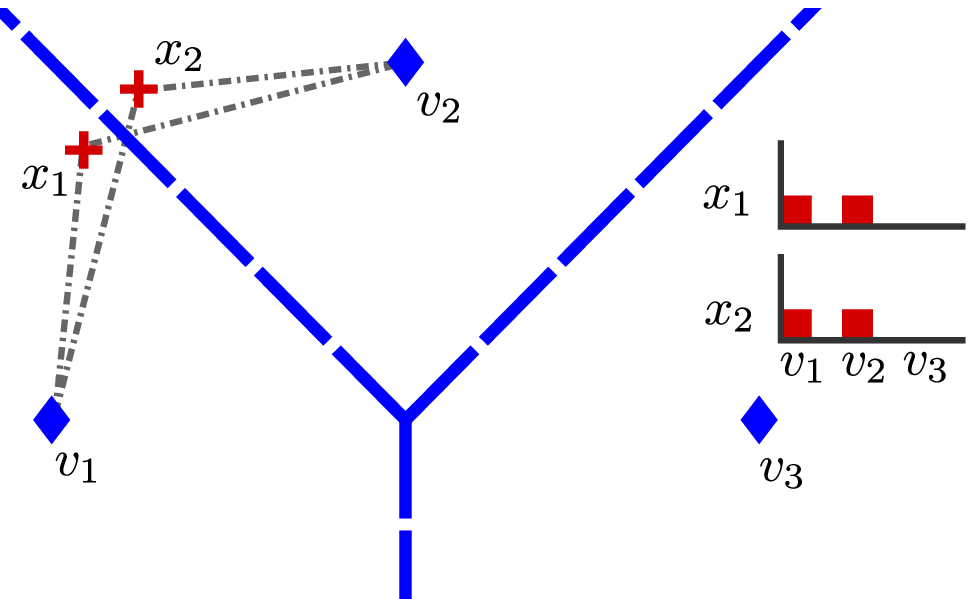}\label{fig:softvq:soft}%
\caption{Two close data points $x_1,x_2$ (+) and the Voronoi partition for three VQ codewords $v_1,v_2,v_3$ ($\blacklozenge$). 
Left: hard VQ ($\tau=1$) assigns similar data points to dissimilar histograms.  
Right: assigning each data point to its top $\tau=2$ codewords reduces noise in codeword histogram representations.\label{fig:softvq}}
\end{figure}

Equation~\eqref{eq:vq} derives from the standard notion of vector quantization (VQ), where each vector (\eg, data point $x_i$) is replaced by its closest quantizer.  However, VQ 
can become unstable when a vector has multiple, (approximately) equidistant quantizers (Figure~\ref{fig:softvq}, left), which is more likely to happen as the
size of the codebook increases.
To counteract quantization errors, we generalize Equation~\eqref{eq:vq} to support \emph{multiple} quantizers for each vector.

For a vector $x_i$, a codebook $\V$, and a \emph{quantization threshold} $\tau\in\{1,2,\dots,|\V|\}$, we define the quantization set 
\[
\argmin_{u\in\V}{}^\tau \|x_i - u\| \doteq \left\{u \text{ is a }\tau\text{-nearest neighbor of } x_i\right\}.
\]
The \emph{top-$\tau$} codeword histogram for a song $x$ is then constructed as
\begin{equation}
h^\tau_x[v] = \frac{1}{|x|} \sum_{x_i\in x} \frac{1}{\tau} \I\left[v \in \argmin_{u\in\V}{}^\tau\|x_i - u\|\right].\label{eq:softvq}
\end{equation}
Intuitively, Equation~\eqref{eq:softvq} assigns $1/\tau$ mass to each of the $\tau$ closest codewords for each $x_i\in x$ (Figure~\ref{fig:softvq}, right).  
Note that when $\tau=1$, Equation~\eqref{eq:softvq} reduces to Equation~\eqref{eq:vq}.  The normalization by $1/\tau$ ensures that $\sum_{v}h^\tau_x[v]=1$, so
that for $\tau>1$, $h^\tau_x$ retains its interpretation as a multinomial distribution over $\V$.

\subsection{Histogram representation and distance}
\label{sec:audio:histogram}

After summarizing each song $x$ by a codeword histogram $h_x^\tau$, these histograms may be interpreted as vectors in $\R^{|\V|}$.  Subsequently, for a query
song $q$, retrieval may be performed by ordering $x\in\X$ according to increasing (Euclidean) distance 
$\|h^\tau_q - h^\tau_x\|$. After optimizing $W$ with Algorithm~\ref{alg:mlr}, the same codeword histogram vectors may be used to perform retrieval with respect to the learned metric $\|h^\tau_q - h^\tau_x\|_W$.

However, treating codeword histograms directly as vectors in a Euclidean space ignores the simplical structure of multinomial distributions.  
To better exploit the geometry of codeword histograms, we represent each histogram in a \emph{probability product kernel} (PPK) space~\cite{jebara2004}. Inner products in this space can be computed by evaluating the corresponding \emph{kernel function} $k$. For PPK space, $k$ is defined as:
\begin{equation}
k(h^\tau_q, h^\tau_x) = \sum_{v\in \V} \sqrt{h^\tau_q[v] \cdot h^\tau_x[v]}.\label{eq:ppk}
\end{equation}
The PPK inner product in Equation~\eqref{eq:ppk} is equivalent to the Bhattacharyya coefficient~\cite{bhattacharyya1943} between $h^\tau_q$ and $h^\tau_x$.  
Consequently, distance in PPK space induces the same rankings as Hellinger distance between histograms.

Typically in kernel methods, data is represented implicitly in a (typically high-dimensional) Hilbert space via the $n\times n$ matrix of inner 
products between training points, \ie, the \emph{kernel matrix}~\cite{scholkopf2002}.  This representation enables efficient learning, even when the
dimensionality of the kernel space is much larger than the number of points (\eg, for histogram-intersection kernels~\cite{barla2003}) or infinite (\eg, 
radial basis functions).  The MLR algorithm has been extended to support optimization of distances in such spaces by reformulating the optimization in terms of 
the kernel matrix, and optimizing an $n\times n$ matrix $W\succeq 0$~\cite{galleguillos2011}.  While kernel MLR supports optimization in arbitrary inner product
spaces, it can be difficult to scale up to large training sets (\ie, large $n$), which may require some approximations, \eg, by restricting $W$ to be
diagonal.

However, for the present application, we can exploit the specific structure of the probability product kernel (on histograms) and optimize distances in PPK space
with complexity that depends on $|\V|$ rather than $n$, thereby supporting larger training sets.  Note that PPK enables an \emph{explicit} representation of 
the data according to a simple, coordinate-wise transformation:
\begin{equation}
h^\tau_x[v] \mapsto \sqrt{h^\tau_x[v]},\label{eq:ppkexplicit}
\end{equation}
which, since $k(h^\tau_x, h^\tau_x)=1$ for all $h^\tau_x$, can be interpreted as mapping the $|\V|$-dimensional simplex to the $|\V|$-dimensional unit sphere.
Training data may therefore be represented as a ${|\V|\times n}$ data matrix, rather than the $n\times n$ kernel matrix.  
As a result, we can equivalently apply Equation~\eqref{eq:ppkexplicit} to the data, and learn a $|\V|\times|\V|$ matrix $W$ with Algorithm~\ref{alg:mlr}, 
which is more efficient than using kernel MLR when $|\V|<n$, as is often the case in our experiments.

\section{Experiments}

\begin{figure}
\centering%
\includegraphics[width=0.45\textwidth]{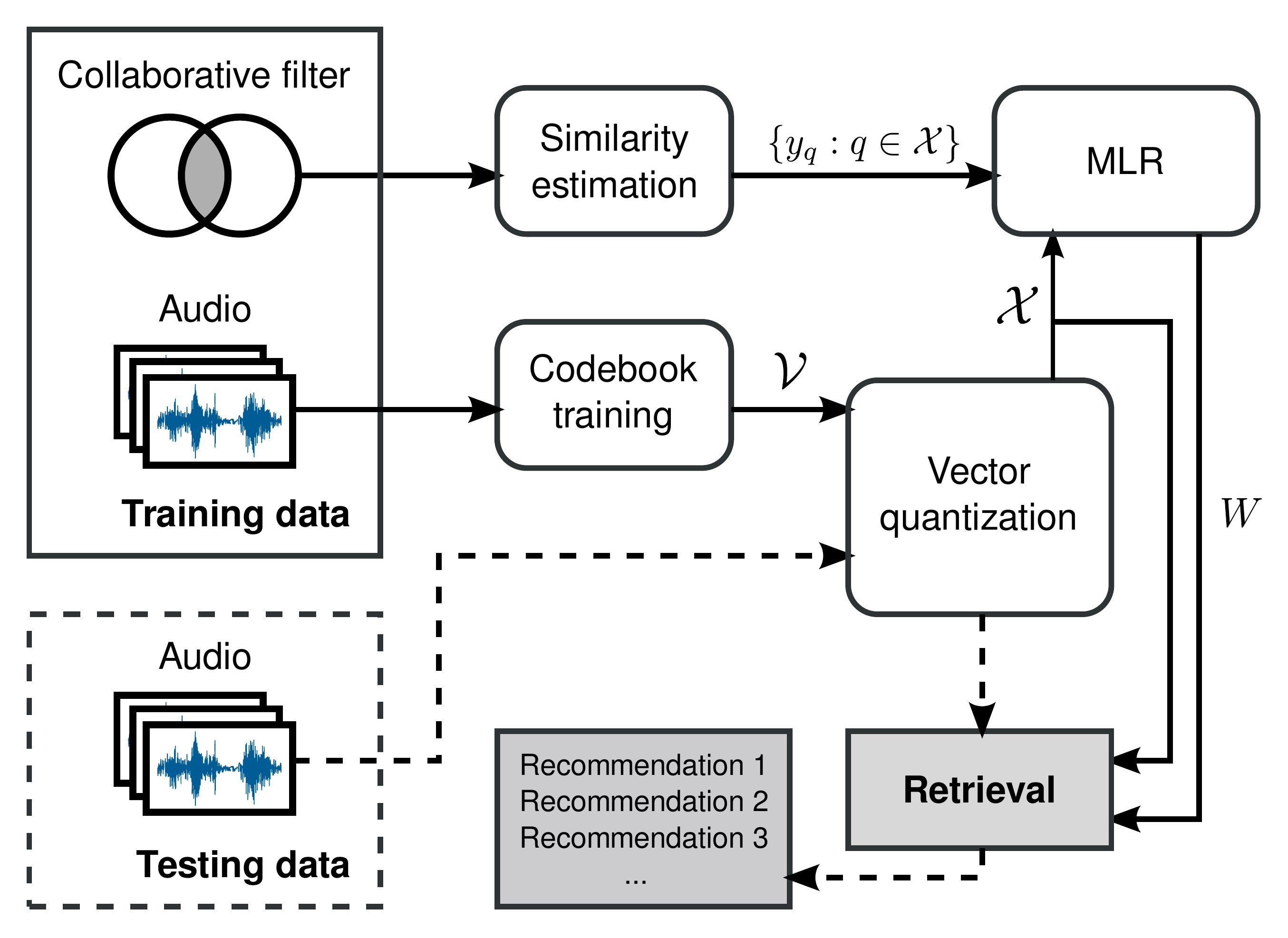}
\caption{Schematic diagram of training and retrieval.  Here, ``training data'' encompasses both the subset of $\X$ used to train the metric $W$, and the codebook set $\X_C$
used to build the codebook $\V$. 
While, in our experiments, both sets are disjoint, in general, data used to build the codebook may also be used to train the metric. 
\label{fig:system}}
\end{figure}

Our experiments are designed to simulate query-by-example content-based retrieval of songs from a fixed database.  Figure~\ref{fig:system} illustrates the high-level experimental setup:
training and evaluation are conducted with respect to collaborative filter similarity (as described in Section~\ref{sec:cfsim}).  In this section, we describe the sources of collaborative 
filter and audio data, experimental procedure, and competing methods against which we compare.

\subsection{Data}
\subsubsection{Collaborative filter: Last.FM}
Our collaborative filter data is provided by Last.fm\footnote{\url{http://www.last.fm/}}, and was collected by Celma~\cite[chapter 3]{celma2010}.  The data
consists of a users-by-artists matrix $F$ of 359,347 unique users and 186,642 unique, identifiable artists; the entry $F[ij]$ contains the number of times user
$i$ listened to artist $j$.  We binarize the matrix by thresholding at 10, \ie, a user must listen to an artist at
least 10 times before we consider the association meaningful.

\vspace{\baselineskip}
\subsubsection{Audio: CAL10K}
For our audio data, we use the CAL10K 
data set~\cite{tingle10}.  Starting from 10,832 songs by 4,661 unique artists, we first partition the set of artists
into those with at least 100 listeners in the binarized CF matrix (2015, the \emph{experiment set}), and those with fewer than 100 listeners (2646, the 
\emph{codebook set}).  We then restrict the CF matrix to just those 2015 artists in the experiment set, with sufficiently many listeners.
From this restricted CF matrix,
we compute the artist-by-artist 
similarity matrix according to Equation~\eqref{eq:jaccard}.

Artists in the codebook set, with insufficiently many listeners, are held out from the experiments in Section~\ref{sec:exp:proc},
but their songs are used to construct four codebooks as described in 
Section~\ref{sec:audio:codebook}.  From each held out artist, we randomly select one song, and extract a 5-second sequence of $\Delta$MFCC vectors (431
half-overlapping 23ms frames at 22050Hz).  These samples are collected into a bag-of-features of approximately 1.1~million 
samples, which is randomly permuted, and clustered via online k-means in a single pass to build four codebooks of sizes $|\V|\in\{256,512,1024,2048\}$, respectively.  Cluster centers are initialized to the first (randomly selected) $k$ points.  
Note that \emph{only} the artists from the codebook set (and thus no artists from the experiment set) are used to construct the codebooks. As a result, the previous four codebooks are fixed throughout the experiments in the following section.

\subsection{Procedure}\label{sec:exp:proc}

For our experiments, we generate 10 random splits of the experiment set of 2015 artists into 40\% training, 30\% validation and 30\% test \emph{artists}\footnote{Due to
recording effects and our definition of similarity, it is crucial to split at the level of artists rather than songs~\cite{whitman2001}.}. For each split, the set of all training artist songs forms the \emph{training set}, which serves as the database of ``known'' songs, $\X$.
For each split, 
and for each (training/test/validation) artist, we then define the \emph{relevant artist set} as the top 10 most similar \emph{training}\footnote{Also for test and validation artists, we restrict the relevant artist set to the training artists to mimic the realistic setting of retrieving ``known'' songs from $\X$, given an ``unknown'' (test/validation) query.} artists. 
Finally, for any song $q$ by artist $i$, we define $q$'s \emph{relevant song set}, $\X_q^+$, as all songs by all artists in $i$'s relevant artist set.
The songs by all other training artists, not in $i$'s relevant artist set, are collected into $\X_q^-$, the set of irrelevant songs for $q$. The statistics of the 
training, validation, and test splits are collected in Table~\ref{tab:swat10kstats}.

\begin{table}
\centering%
\begin{tabular}{lrrr}
            & Training          & Validation        & Test\\
 \hline
\# Artists 	& 806	            & 604	            & 605\\
\# Songs    & 2122.3 $\pm$ 36.3	& 1589.3 $\pm$ 38.6	& 1607.5 $\pm$ 64.3\\
\# Relevant	& 36.9   $\pm$ 16.4	& 36.4   $\pm$ 15.4	& 37.1   $\pm$ 16.0
\end{tabular}
\caption{Statistics of CAL10K data, averaged across ten random training/validation/test splits.  
\emph{\# Relevant} is the average number of relevant songs for each training/validation/test song.\label{tab:swat10kstats}}
\end{table}

For each of the four codebooks, constructed in the previous section, each song was represented by a histogram over codewords using Equation~\eqref{eq:softvq}, with $\tau\in\{1,2,4,8\}$.  Codeword histograms were
then mapped into PPK space by Equation~\eqref{eq:ppkexplicit}.  For comparison purposes, we also experiment with Euclidean distance and MLR on the raw codeword
histograms.

To train the distance metric with Algorithm~\ref{alg:mlr}, we vary $C\in\{10^{-2}, 10^{-1}, \cdots, 10^9\}$.  We experiment with three ranking losses $\Delta$ 
for training: area under the ROC curve (AUC), which captures global qualities of the ranking, but penalizes mistakes equally regardless of their position in the ranking; normalized discounted cumulative gain (NDCG), which applies larger penalties to mistakes at the beginning of the ranking than at the end, and is 
therefore more localized than AUC; and mean reciprocal rank (MRR), which is determined by the position of the first relevant result, and is therefore the 
most localized ranking loss under consideration here. 
After learning $W$ on the training set, retrieval is evaluated on the validation set, and the parameter 
setting $(C,\Delta)$ which achieves highest AUC on the validation set is then evaluated on the test set.

To evaluate a metric $W$, the training set $\X$ is ranked according to distance from each test (validation) song $q$ under $W$, and we record the mean AUC of
the rankings over all test (validation) songs.

Prior to training with MLR, codeword histograms are compressed via principal components analysis (PCA) to capture 95\% of the variance as estimated 
on the training set.  While primarily done for computational efficiency, this step is similar to the latent perceptual indexing method described by Sundaram and
Narayanan~\cite{sundaram2008}, and may also be interpreted as de-noising the codeword histogram representations.  In preliminary experiments, compression of
codeword histograms was not observed to significantly affect retrieval accuracy in either the native or PPK spaces (without MLR optimization). 

\subsection{Comparisons}
\label{sec:comparisons}
To evaluate the performance of the proposed system, we compare to several alternative methods for content-based
query-by-example song retrieval: first, similarity derived from comparing Gaussian mixture models of $\Delta$MFCCs; second, an alternative (unsupervised) weighting of VQ codewords; and third, a high-level, automatic semantic annotation method.  We also include a comparison to a manual semantic annotation method (\ie, driven by human experts), which although not content-based, can provide an estimate of an upper bound on what can be achieved by content-based methods.
For both manual and automatic semantic annotations, we will also compare to their MLR-optimized counterparts.

\vspace{\baselineskip}
\subsubsection{Gaussian mixtures}

From each song, a Gaussian mixture model (GMM) over its $\Delta$MFCCs was estimated via 
expectation-maximization~\cite{dempster1977}. 
Each GMM consists of 8 components with diagonal covariance.  The training set 
$\X$ is therefore
represented as a collection of GMM distributions ${\{p_x:~x\in\X\}}$.  This approach is fairly standard in music
information retrieval~\cite{aucouturier02,berenzweig2004,jensen2007}, and is intended to serve as a baseline against which we can compare the
proposed VQ approach.

At test time, given a query song $q$, we first estimate its GMM $p_q$.  We would then like to rank 
each $x\in\X$ by increasing Kullback-Leibler (KL) divergence~\cite{kullback1968} from $p_q$:
\begin{equation}
D(p_q \| p_x) = \int p_q(z) \log\frac{p_q(z)}{p_x(z)}\mathrm{d}z.\label{eq:kl}
\end{equation}
However, we do not have a closed-form expression for KL~divergence between GMMs, so we must
resort to approximate methods.  Several such approximation schemes have been devised in recent years, including 
variational methods and sampling approaches~\cite{hershey2007}.  Here, we opt for the Monte Carlo approximation:
\begin{equation}
D(p_q \| p_x) \approx \sum_{i=1}^{m} \frac{1}{m} \log\frac{p_q(z_i)}{p_x(z_i)},\label{eq:mckl}
\end{equation}
where $\{z_i\}_{i=1}^m$ is a collection of $m$ independent samples drawn from $p_q$. 
Although the Monte Carlo approximation is 
considerably slower than closed-form approximations (\eg, variational methods), with enough samples, it often exhibits higher 
accuracy~\cite{hershey2007,jensen2007}.
Note that because we are only interested in the rank-ordering of $\X$ given $p_q$, it is equivalent to order each $p_x\in\X$ by 
increasing (approximate) cross-entropy:
\begin{equation}
H(p_q,p_x) = \int p_q(z) \log \frac{1}{p_x(z)}\mathrm{d}z \approx \sum_{i=1}^m \frac{1}{m} \log \frac{1}{p_x(z_i)}.\label{eq:mcentropy}
\end{equation}
For efficiency purposes, for each query $q$ we fix the sample $\{z_i\}_{i=1}^m \sim p_q$ across all $x\in\X$.  We use
$m=2048$ samples for each query, which was found to yield stable cross-entropy estimates in an informal,
preliminary experiment.

\vspace{\baselineskip}
\subsubsection{TF-IDF}

The algorithm described in Section~\ref{sec:learning:mlr} is a supervised approach to learning an optimal 
transformation of feature descriptors (in this specific case, VQ histograms).  Alternatively, it is common to use 
the natural statistics of the data in an unsupervised fashion to transform the feature descriptors.  As a baseline, we
compare to the standard method of combining \emph{term frequency}--\emph{inverse document frequency} (TF-IDF)~\cite{tfidf} 
representations with cosine similarity, which is commonly used with both text~\cite{tfidf} and codeword representations~\cite{sivic2003}.

Given a codeword histogram $h^\tau_q$, for each $v\in\V$, $h^\tau_q[v]$ is mapped to its TF-IDF value by\footnote{Since codeword histograms are pre-normalized, there is no need to re-compute the term frequency in Equation~\eqref{eq:tfidf}.}
\begin{equation}
h^\tau_q[v] \mapsto h^\tau_q[v] \cdot \text{IDF}[v],\label{eq:tfidf}
\end{equation}
where IDF[$v$] is computed from the statistics of the training set by\footnote{To avoid division by 0, we define $\text{IDF}[v]=0$ for any codeword $v$ which is not used in the training set.}
\begin{equation}
\text{IDF}[v] = \log\frac{|\X|}{|\{x \in \X:~ x[v] > 0\}|}.\label{eq:idf}
\end{equation}
Intuitively, $\text{IDF}[v]$ assigns more weight to codewords
$v$ which appear in fewer songs, and reduces the importance of codewords appearing in many songs.
The training set $\X$ 
is
accordingly represented by TF-IDF vectors.  At test time, each $x\in\X$ is ranked according to decreasing 
cosine-similarity to the query $q$:
\begin{equation}
\cos(h^\tau_q,h^\tau_x) = \frac{{h^\tau_q}\trans h^\tau_x}{\|h^\tau_q\|\cdot\|h^\tau_x\|}.\label{eq:cosine}
\end{equation}

\vspace{\baselineskip}
\subsubsection{Automatic semantic tags}
The proposed method relies on low-level descriptors to assess similarity between songs.  Alternatively,
similarity may be assessed by comparing high-level content descriptors in the form of \emph{semantic tags}.  
These tags may include words to describe genre, instrumentation, emotion, \etc.  Because semantic annotations may not
be available for novel query songs, we restrict attention to algorithms which automatically predict tags given 
only audio content.

In our experiments, we adapt the auto-tagging method proposed by Turnbull~\etal.~\cite{turnbull2008}.  This method summarizes each song
by a \emph{semantic multinomial distribution} (SMD) over a vocabulary of 149 tag words.  Each tag $t$ is characterized 
by a GMM $p_t$ over $\Delta$MFCC vectors, each of which was trained previously on the CAL500 data set~\cite{turnbull2007}.  A song $q$ is
summarized by a multinomial distribution $s_q$, 
where the $t\th$ entry 
is computed by the geometric mean of the likelihood of
$q$'s $\Delta$MFCC vectors $q_i$ under $p_t$:
\begin{equation}
s_q[t] \propto \left(\prod_{q_i \in q} p_t(q_i)\right)^{1/|q|}.\label{eq:smd}
\end{equation}
(Each SMD 
$s_q$ is normalized to sum to 1.)
The training set 
$\X$ is thus described as a collection of SMDs $\{s_x:~ x \in \X\}$.  At test time, $\X$
is ranked according to increasing distance from the test query under 
the probability product kernel\footnote{We also experimented with $\chi^2$-distance, $\ell_1$, Euclidean, and (symmetrized) KL divergence, but PPK distance was
always statistically equivalent to the best-performing distance.}
as described in Section~\ref{sec:audio:histogram}.  This representation is also amenable to optimization by MLR, and we will compare to retrieval performance
after optimizing PPK representations of SMDs with MLR.

\vspace{\baselineskip}
\subsubsection{Human tags}
Our final comparison uses semantic annotations manually produced by humans, and may therefore be interpreted as an
upper bound on how well we may expect content-based methods to perform.  Each song in CAL10K includes a partially
observed, binary annotation vector over a vocabulary of 1053 tags from the Music Genome
Project\footnote{\url{http://www.pandora.com/mgp.shtml}}.  The annotation vectors are \emph{weak} in the sense that a 1
indicates that the tag applies, while a 0 indicates only that the tag \emph{may not} apply.  

In our experiments, we observed the best performance by using cosine similarity as the retrieval function, although we
also tested TF-IDF and Euclidean distances.  As in the auto-tag case, we will also compare to tag vectors after 
optimization by MLR. 
When training with MLR, annotation vectors were compressed via PCA to capture 95\% of the training set variance.

\section{Results}

\subsection*{Vector quantization}

\begin{figure}
\centering%
\includegraphics[width=0.5\textwidth]{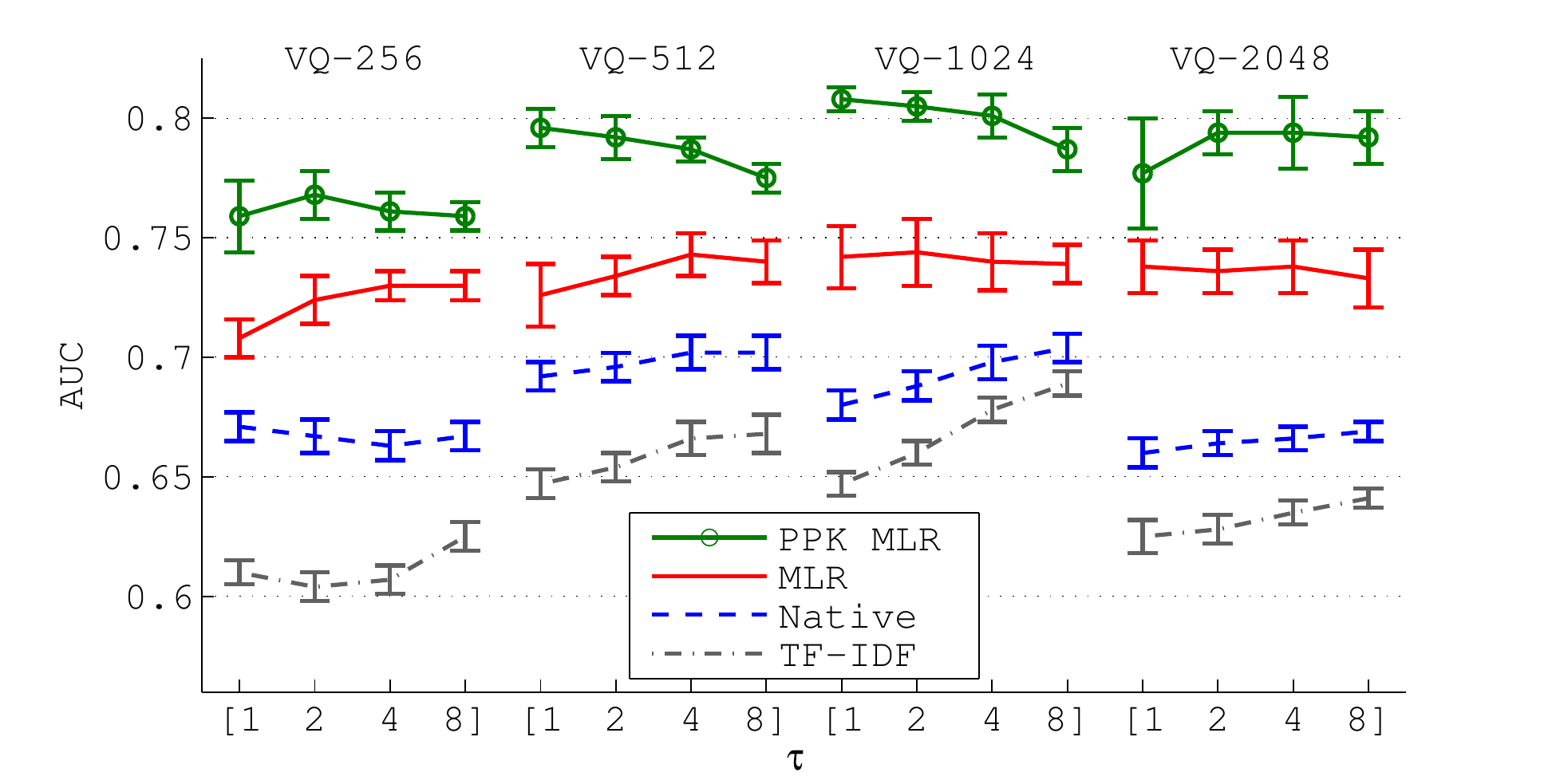}
\caption{
Retrieval accuracy with vector quantized $\Delta$MFCC representations.  Each grouping corresponds to a different
codebook size $|\V|\in\{256,512,1024,2048\}$.  Each point within a group corresponds to a different quantization threshold $\tau\in\{1,2,4,8\}$.
\emph{TF-IDF} refers to cosine similarity applied to IDF-weighted VQ histograms; \emph{Native} refers to Euclidean distance on unweighted VQ histograms; \emph{MLR} refers to VQ histograms after optimization by MLR; \emph{PPK MLR} refers to distances after mapping VQ histograms into probability product kernel space and subsequently optimizing with MLR.
Error bars correspond to one standard deviation across trials.  
\label{fig:vqresults}}
\end{figure}

In a first series of experiments, we evaluate various approaches and configurations based on VQ codeword histograms. Figure~\ref{fig:vqresults} lists the AUC achieved by four different approaches (\emph{Native}, \emph{TF-IDF}, \emph{MLR}, \emph{PPK-MLR}), based on VQ codeword histograms, for each of four codebook sizes and each of four quantization thresholds.
We observe that using Euclidean distance on raw codeword histograms\footnote{For clarity, we omit the performance curves for 
native Euclidean distance on PPK representations, as they do not differ significantly from the \emph{Native} curves shown.}
(\emph{Native}) yields significantly higher performance for codebooks of intermediate size (512 or 1024) than for small (256) or large (2048) codebooks.  
For the 1024 codebook, increasing $\tau$ results in significant gains in performance, but it does not exceed the performance for the 512 codebook.  
The decrease in accuracy for $|\V|=2048$ suggests that performance is indeed sensitive to overly large codebooks.

After learning an optimal distance metric with MLR on raw histograms (\ie, not PPK representations) (\emph{MLR}), we observe two interesting effects.  
First, MLR optimization always yields significantly better performance than the native Euclidean distance.  
Second, performance is much less sensitive to the choice of codebook size and quantization threshold: all settings of $\tau$ for codebooks of size at least 
$|\V|\geq512$ achieve statistically equivalent performance.

Finally, we observe the highest performance by combining the PPK representation with MLR optimization (\emph{PPK-MLR}).  For ${|\V|=1024,\tau=1}$, the mean AUC score improves 
from $0.680\pm0.006$ (Native) to $0.808\pm0.005$ (PPK-MLR).  The effects of codebook size and quantization threshold are diminished by MLR optimization, 
although they are slightly more pronounced than in the previous case without PPK. We may then ask: does top-$\tau$ VQ provide any benefit? 

\begin{figure}
\centering%
\includegraphics[width=0.5\textwidth]{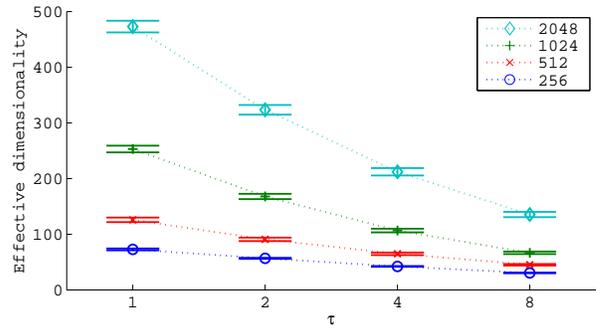}
\caption{
The \emph{effective dimensionality} 
of codeword histograms in PPK space, \ie, the number of principal components necessary to capture 95\% of the training set's variance, as a function of the quantization threshold $\tau$.
(The results reported in the figure 
are the average effective dimension $\pm$ one standard deviation across trials.)
\label{fig:ppkdimension}}
\end{figure}

Figure~\ref{fig:ppkdimension} lists the effective dimensionality --- the number of principal components necessary to capture 95\% of the training set's variance --- 
of codeword histograms in PPK space as a function of quantization threshold $\tau$.
Although for the best-performing codebook size $|\V|=1024$, each of $\tau\in\{1,2,4\}$ achieves statistically equivalent performance,
the effective dimensionality varies from $253.1\pm6.0$ ($\tau=1$) to $106.6\pm3.3$ ($\tau=4$).  
Thus, top-$\tau$ VQ can be applied to dramatically reduce the dimensionality of VQ representations, which in turn reduces the number of parameters learned by
MLR, and therefore improves the efficiency 
of learning and retrieval, without significantly degrading performance.

\subsection*{Qualitative results}
\begin{figure*}[!t]
\centering%
\includegraphics[width=\textwidth]{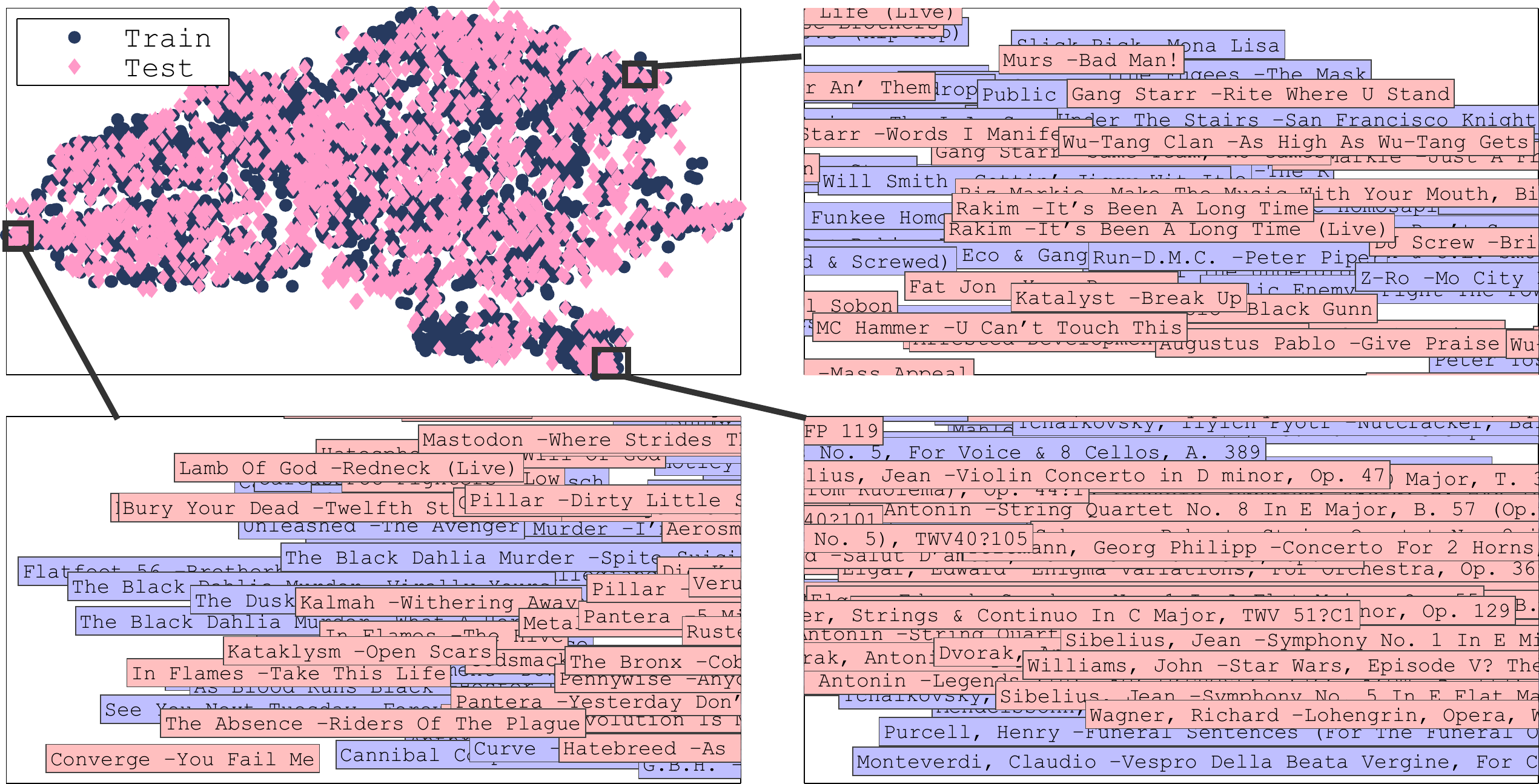}%
\caption{
A t-SNE visualization of the optimized similarity space produced by PPK+MLR 
on one training/test split of the data ($|\V|=1024$, $\tau=1$).
Close-ups on three peripheral regions reveal \emph{hip-hop} (upper-right), \emph{metal} (lower-left), and \emph{classical} (lower-right) 
genres.\label{fig:scatterplot}}
\end{figure*}

Figure~\ref{fig:scatterplot} illustrates an example optimized similarity space produced by MLR on PPK histogram representations,
as visualized in two
dimensions by t-SNE~\cite{tsne}.  Even though the algorithm is never exposed to any explicit semantic information, the 
optimized space does exhibit regions which seem to capture intuitive notions of genre, such as \emph{hip-hop}, \emph{metal},
and \emph{classical}.

Table~\ref{tab:playlist} illustrates a few example queries and their top-5 closest results under the Euclidean and 
MLR-optimized metric.  The native space seems to capture similarities due to energy and instrumentation, but does 
not necessarily match CF similarity.  The optimized space captures aspects of the audio data which correspond to
CF similarity, and produces playlists with more relevant results. 
\begin{table*}
\centering%
\begin{tabular}{p{0.26\textwidth}r@{}p{0.32\textwidth}r@{}p{0.32\textwidth}}
Test query                                                              && VQ (Native)           && VQ (PPK+MLR)\\
\hline
\multirow{5}{*}{\parbox{0.26\textwidth}{Ornette Coleman - Africa is the Mirror of All Colors}}
    &&Judas Priest - You've Got Another Thing Comin'  &  &Wynton Marsalis - Caravan                  \\
    &&Def Leppard - Rock of Ages                      &  $\blacktriangleright$& Dizzy Gillespie - Dizzy's Blues\\
    &&KC \& The Sunshine Band - Give it Up            &  $\blacktriangleright$& Michael Brecker - Two Blocks from the Edge\\
    &&Wynton Marsalis - Caravan                       &  $\blacktriangleright$& Eric Dolphy - Miss Ann (live)\\
    &&Ringo Starr - It Don't Come Easy                &&  Ramsey Lewis - Here Comes Santa Claus\\
\hline
\multirow{5}{*}{Fats Waller - Winter Weather}
    &$\blacktriangleright$& Dizzy Gillespie - She's Funny that Way    &&Chet Atkins - In the Mood \\                                                           
    &&Enrique Morente - Solea                                         &$\blacktriangleright$& Charlie Parker - What Is This Thing Called Love? \\
    &&Chet Atkins - In the Mood                                       &$\blacktriangleright$& Bud Powell - Oblivion \\
    &&Rachmaninov - Piano Concerto \#4 in Gmin                        &$\blacktriangleright$& Bob Wills \& His Texas Playboys - Lyla Lou \\
    &&Eluvium - Radio Ballet                                          &$\blacktriangleright$& Bob Wills \& His Texas Playboys - Sittin' On Top Of The World \\
\hline
\multirow{5}{*}{The Ramones - Go Mental}
    && Def Leppard - Promises                   &$\blacktriangleright$&  The Buzzcocks - Harmony In My Head\\ 
    &$\blacktriangleright$&  The Buzzcocks - Harmony In My Head       && Motley Crue - Same Ol' Situation\\   
    && Los Lonely Boys - Roses                  &$\blacktriangleright$&  The Offspring - Gotta Get Away\\     
    && Wolfmother - Colossal                    &$\blacktriangleright$&  The Misfits - Skulls\\               
    && Judas Priest - Diamonds and Rust (live)  &$\blacktriangleright$&  AC/DC - Who Made Who (live)\\
\hline
\end{tabular}
\caption{Example playlists generated by 5-nearest (training) neighbors of three different query (test) songs (left) using Euclidean distance on raw codeword
histograms (center) and MLR-optimized PPK distances (right).
Relevant results are indicated by $\blacktriangleright$.\label{tab:playlist}}
\end{table*}

\subsection*{Comparison}
\begin{figure}
\centering%
\includegraphics[width=0.5\textwidth]{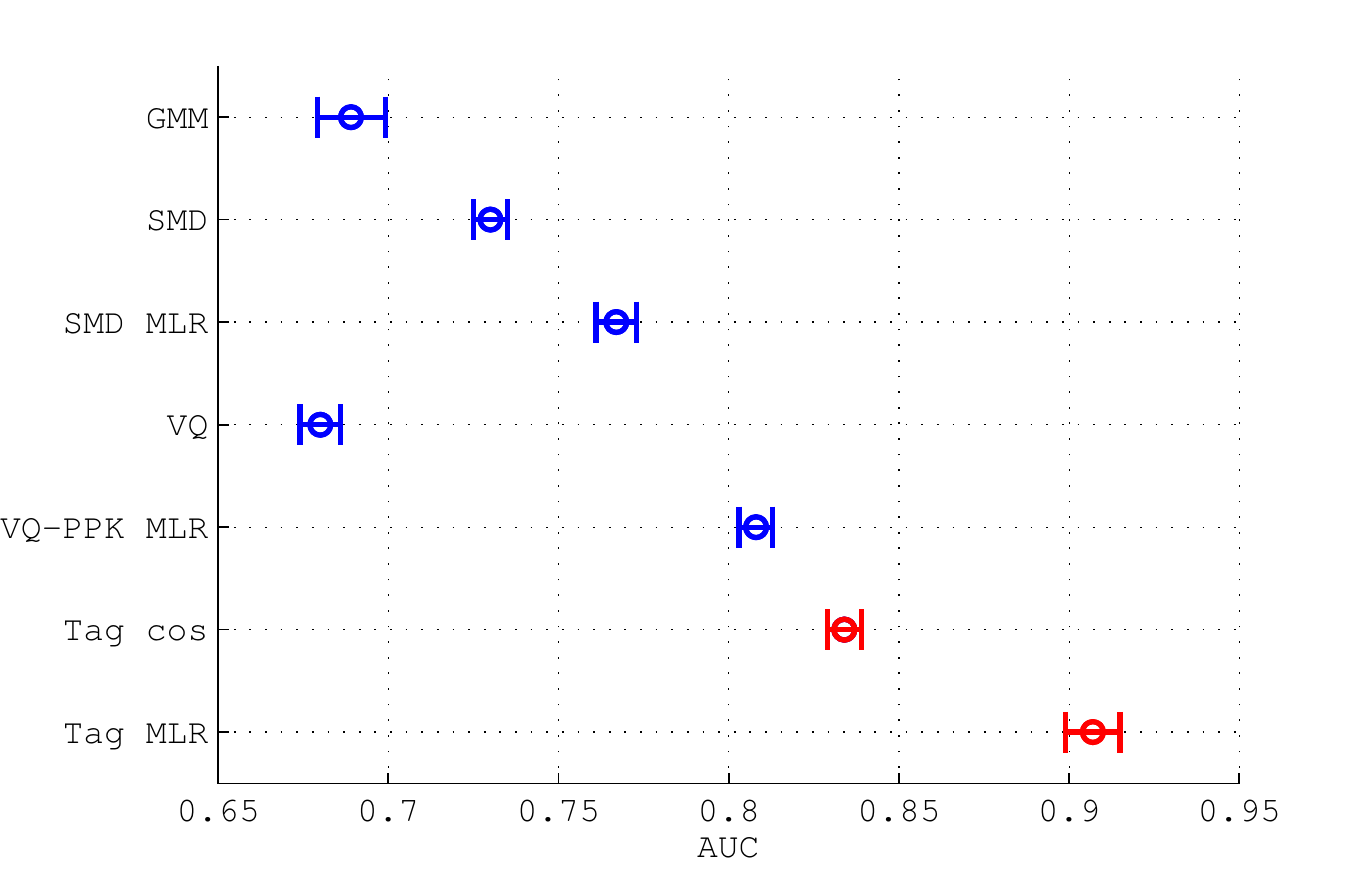}
\caption{Comparison of VQ-based retrieval accuracy to competing methods.  \emph{VQ} corresponds to a codebook of size $V=1024$ with
quantization threshold $\tau=1$.  \emph{Tag}-based methods (red) use human annotations, and are not automatically derived from audio content.
Error bars correspond to one standard deviation across trials.\label{fig:comparison}}
\end{figure}

Figure~\ref{fig:vqresults} lists the accuracy achieved by using TF-IDF weighting on codeword histograms.  For all VQ configurations (\ie, for each codebook size and quantization threshold) TF-IDF significantly
degrades performance compared to MLR-based methods, which indicates that inverse document frequency may not be as an accurate predictor of salience in codeword histograms as in natural
language~\cite{tfidf}.

Figure~\ref{fig:comparison} shows the performance of all other methods against which we compare.  First, we observe that \emph{raw} SMD representations 
provide more accurate retrieval than both the GMM approach and \emph{raw} VQ codeword histograms (\ie, prior to optimization by MLR).  This may be
expected, as previous studies have demonstrated superior query-by-example retrieval performance when using semantic representations of multimedia data~\cite{rasiwasia2007,barrington2007}.

Moreover, SMD and VQ can be optimized by MLR to achieve significantly higher performance than raw SMD and VQ, respectively. The semantic representations in SMD compress the original audio content to a small set of descriptive terms, at a higher level of abstraction. In raw form, this representation provides a more robust set of features, which improves recommendation performance compared to matching low-level content features that are often noisier. On the other hand, semantic representations are inherently limited by the choice of vocabulary and may prematurely discard important discriminative information (\eg, subtle distinctions within sub-genres). This renders them less attractive as starting point for a metric learning algorithm like MLR, compared to less-compressed (but possibly noisier) representations, like VQ. Indeed, the latter may retain more information for MLR to learn an appropriate similarity function.
This is confirmed by our experiments: MLR improves VQ significantly more than it does for SMD. As a result, MLR-VQ outperforms all other content-based methods in our experiments.

Finally, we provide an estimate of an upper bound on what can be achieved by automatic, content-based methods, by evaluating the retrieval performance when using manual annotations (\emph{Tag} in Figure~\ref{fig:comparison}):
$0.834\pm0.005$ with cosine similarity, and $0.907\pm0.008$ with MLR-optimized similarity.  The
improvement in accuracy for human tags, when using MLR, indicates that even hand-crafted annotations can be improved by learning an optimal distance over tag vectors.  
By
contrast, TF-IDF on human tag vectors decreases performance to $0.771\pm0.004$, indicating that IDF does not accurately model (binary) tag salience.  The gap in
performance between content-based methods and manual annotations suggests that there is still room for improvement.  Closing this gap may require
incorporating more complex features to capture rhythmic and structural properties of music which are discarded by the simple timbral descriptors used here.

\section{Conclusion}

In this article, we have proposed a method for improving content-based audio similarity by learning from a sample of collaborative filter data.  Collaborative
filters form the basis of state-of-the-art recommendation systems, but cannot directly form recommendations or answer queries for items which have not yet been
consumed or rated.  By optimizing content-based similarity from a collaborative filter, we provide a simple mechanism for alleviating the cold-start problem and
extending music recommendation to novel or less known songs.

By using implicit feedback in the form of user listening history, we can efficiently collect high-quality training data without active user participation, 
and as a result, train on larger collections of music than would be practical with explicit feedback or survey data. 
Our notion of similarity derives from user activity in a bottom-up fashion, and obviates the need for
coarse simplifications such as genre or artist agreement.  

Our proposed top-$\tau$ VQ audio representation enables efficient and compact description of the acoustic content of music data.  Combining this audio representation
with an optimized distance metric yields similarity calculations which are both efficient to compute and substantially more accurate than competing 
content-based methods.

While in this work, our focus remains on music recommendation applications, the proposed methods are quite general, and may apply to a wide variety of
applications involving content-based similarity, such as nearest-neighbor classification of audio signals.


%



\section*{Acknowledgment}

The authors acknowledge support from Qualcomm, Inc, eHarmony, Inc., Yahoo! Inc., and NSF Grants CCF-0830535 and IIS-1054960.  This research was supported in part by the UCSD FWGrid Project, NSF Research Infrastructure Grant Number~EIA-0303622.

\ifCLASSOPTIONcaptionsoff
  \newpage
\fi



\bibliographystyle{IEEEtran}
\bibliography{refs}

\begin{thebibliography}{10}
\providecommand{\url}[1]{#1}
\csname url@samestyle\endcsname
\providecommand{\newblock}{\relax}
\providecommand{\bibinfo}[2]{#2}
\providecommand{\BIBentrySTDinterwordspacing}{\spaceskip=0pt\relax}
\providecommand{\BIBentryALTinterwordstretchfactor}{4}
\providecommand{\BIBentryALTinterwordspacing}{\spaceskip=\fontdimen2\font plus
\BIBentryALTinterwordstretchfactor\fontdimen3\font minus
  \fontdimen4\font\relax}
\providecommand{\BIBforeignlanguage}[2]{{%
\expandafter\ifx\csname l@#1\endcsname\relax
\typeout{** WARNING: IEEEtran.bst: No hyphenation pattern has been}%
\typeout{** loaded for the language `#1'. Using the pattern for}%
\typeout{** the default language instead.}%
\else
\language=\csname l@#1\endcsname
\fi
#2}}
\providecommand{\BIBdecl}{\relax}
\BIBdecl

\bibitem{goldberg1992}
\BIBentryALTinterwordspacing
D.~Goldberg, D.~Nichols, B.~M. Oki, and D.~Terry, ``Using collaborative
  filtering to weave an information tapestry,'' \emph{Commun. ACM}, vol.~35,
  pp. 61--70, December 1992. [Online]. Available:
  \url{http://doi.acm.org/10.1145/138859.138867}
\BIBentrySTDinterwordspacing

\bibitem{genius}
L.~Barrington, R.~Oda, and G.~Lanckriet, ``Smarter than genius? {H}uman
  evaluation of music recommender systems,'' in \emph{Proceedings of the 10th
  International Conference on Music Information Retrieval}, 2009.

\bibitem{kim09}
J.~H. Kim, B.~Tomasik, and D.~Turnbull, ``Using artist similarity to propagate
  semantic information,'' in \emph{Proceedings of the 10th International
  Conference on Music Information Retrieval}, 2009.

\bibitem{celma2010}
O.~Celma, \emph{{Music Recommendation and Discovery in the Long Tail}}.\hskip
  1em plus 0.5em minus 0.4em\relax Springer, 2010.

\bibitem{logan2001}
B.~Logan and A.~Salomon, ``A music similarity function based on signal
  analysis,'' \emph{Multimedia and Expo, IEEE International Conference on},
  vol.~0, p. 190, 2001.

\bibitem{aucouturier02}
J.-J. Aucouturier and F.~Pachet, ``Music similarity measures: What's the use?''
  in \emph{Inernational Symposium on Music Information Retrieval (ISMIR2002)},
  2002, pp. 157--163.

\bibitem{ellis02}
D.~Ellis, B.~Whitman, A.~Berenzweig, and S.~Lawrence, ``The quest for ground
  truth in musical artist similarity,'' in \emph{Proeedings of the
  International Symposium on Music Information Retrieval (ISMIR)}, October
  2002, pp. 170--177.

\bibitem{berenzweig2004}
A.~Berenzweig, B.~Logan, D.~P. Ellis, and B.~Whitman, ``A large-scale
  evaluation of acoustic and subjective music-similarity measures,''
  \emph{Computer Music Journal}, vol.~28, no.~2, pp. 63--76, 2004.

\bibitem{slaney08}
M.~Slaney, K.~Weinberger, and W.~White, ``Learning a metric for music
  similarity,'' in \emph{International Symposium on Music Information Retrieval
  (ISMIR2008)}, September 2008, pp. 313--318.

\bibitem{mcfee2011}
B.~{McFee} and G.~Lanckriet, ``Learning multi-modal similarity,'' \emph{Journal
  of Machine Learning Research}, vol.~12, pp. 491--523, February 2011.

\bibitem{slaney07}
M.~Slaney and W.~White, ``Similarity based on rating data,'' in
  \emph{International Symposium on Music Information Retrieval (ISMIR2007)},
  2007, pp. 479--484.

\bibitem{yoshii08}
K.~Yoshii, M.~Goto, K.~Komatani, T.~Ogata, and H.~Okuno, ``An efficient hybrid
  music recommender system using an incrementally trainable probabilistic
  generative model,'' \emph{IEEE Transactions on Audio, Speech, and Language
  Processing}, vol.~16, no.~2, pp. 435--447, 2008.

\bibitem{stenzel2005}
R.~Stenzel and T.~Kamps, ``Improving content-based similarity measures by
  training a collaborative model,'' in \emph{International Symposium on Music
  Information Retrieval (ISMIR2005)}, 2005, pp. 264--271.

\bibitem{mcfee10a}
B.~McFee and G.~Lanckriet, ``Metric learning to rank,'' in \emph{Proceedings of
  the 27th annual International Conference on Machine Learning (ICML)},
  J.~F{\"u}rnkranz and T.~Joachims, Eds., Haifa, Israel, June 2010, pp.
  775--782.

\bibitem{sarwar2001}
\BIBentryALTinterwordspacing
B.~Sarwar, G.~Karypis, J.~Konstan, and J.~Reidl, ``Item-based collaborative
  filtering recommendation algorithms,'' in \emph{Proceedings of the 10th
  international conference on World Wide Web}, ser. WWW '01.\hskip 1em plus
  0.5em minus 0.4em\relax New York, NY, USA: ACM, 2001, pp. 285--295. [Online].
  Available: \url{http://doi.acm.org/10.1145/371920.372071}
\BIBentrySTDinterwordspacing

\bibitem{deshpande2004}
\BIBentryALTinterwordspacing
M.~Deshpande and G.~Karypis, ``Item-based top-n recommendation algorithms,''
  \emph{ACM Trans. Inf. Syst.}, vol.~22, pp. 143--177, January 2004. [Online].
  Available: \url{http://doi.acm.org/10.1145/963770.963776}
\BIBentrySTDinterwordspacing

\bibitem{hu2008}
Y.~Hu, Y.~Koren, and C.~Volinsky, ``Collaborative filtering for implicit
  feedback datasets,'' in \emph{Data Mining, 2008. ICDM '08. Eighth IEEE
  International Conference on}, 2008, pp. 263 --272.

\bibitem{jaccard1901}
P.~Jaccard, ``{\'{E}tude comparative de la distribution florale dans une
  portion des Alpes et des Jura},'' \emph{Bulletin del la Soci\'{e}t\'{e}
  Vaudoise des Sciences Naturelles}, vol.~37, pp. 547--579, 1901.

\bibitem{xing2003}
E.~P. Xing, A.~Y. Ng, M.~I. Jordan, and S.~Russell, ``Distance metric learning,
  with application to clustering with side-information,'' in \emph{Advances in
  Neural Information Processing Systems 15}.\hskip 1em plus 0.5em minus
  0.4em\relax Cambridge, MA: MIT Press, 2003, pp. 505--512.

\bibitem{weinberger2006}
K.~Q. Weinberger, J.~Blitzer, and L.~K. Saul, ``Distance metric learning for
  large margin nearest neighbor classification,'' in \emph{Advances in Neural
  Information Processing Systems 18}, Y.~Weiss, B.~Sch\"{o}lkopf, and J.~Platt,
  Eds.\hskip 1em plus 0.5em minus 0.4em\relax Cambridge, MA: MIT Press, 2006,
  pp. 451--458.

\bibitem{davis2007}
\BIBentryALTinterwordspacing
J.~V. Davis, B.~Kulis, P.~Jain, S.~Sra, and I.~S. Dhillon,
  ``Information-theoretic metric learning,'' in \emph{Proceedings of the 24th
  international conference on Machine learning}, ser. ICML '07.\hskip 1em plus
  0.5em minus 0.4em\relax New York, NY, USA: ACM, 2007, pp. 209--216. [Online].
  Available: \url{http://doi.acm.org/10.1145/1273496.1273523}
\BIBentrySTDinterwordspacing

\bibitem{egan1975}
J.~P. Egan, \emph{Signal detection theory and {ROC} analysis}, ser. Series in
  Cognition and Perception.\hskip 1em plus 0.5em minus 0.4em\relax New York,
  NY: Academic Press, 1975.

\bibitem{tsochantaridis06}
I.~Tsochantaridis, T.~Joachims, T.~Hofmann, and Y.~Altun, ``Large margin
  methods for structured and interdependent output variables,'' \emph{J. Mach.
  Learn. Res.}, vol.~6, pp. 1453--1484, 2005.

\bibitem{joachims2005}
T.~Joachims, ``A support vector method for multivariate performance measures,''
  in \emph{Proceedings of the 22nd international conference on Machine
  learning}.\hskip 1em plus 0.5em minus 0.4em\relax New York, NY, USA: ACM,
  2005, pp. 377--384.

\bibitem{yue2007}
Y.~Yue, T.~Finley, F.~Radlinski, and T.~Joachims, ``A support vector method for
  optimizing average precision,'' in \emph{SIGIR '07: Proceedings of the 30th
  annual international ACM SIGIR conference on Research and development in
  information retrieval}.\hskip 1em plus 0.5em minus 0.4em\relax New York, NY,
  USA: ACM, 2007, pp. 271--278.

\bibitem{chakrabarti2008}
S.~Chakrabarti, R.~Khanna, U.~Sawant, and C.~Bhattacharyya, ``Structured
  learning for non-smooth ranking losses,'' in \emph{KDD '08: Proceeding of the
  14th ACM SIGKDD international conference on Knowledge discovery and data
  mining}.\hskip 1em plus 0.5em minus 0.4em\relax New York, NY, USA: ACM, 2008,
  pp. 88--96.

\bibitem{voorhees2001}
E.~M. Voorhees, ``Overview of the trec 2001 question answering track,'' in
  \emph{In Proceedings of the Tenth Text REtrieval Conference (TREC)}, 2001,
  pp. 42--51.

\bibitem{jarvelin2000}
K.~J\"{a}rvelin and J.~Kek\"{a}l\"{a}inen, ``{IR} evaluation methods for
  retrieving highly relevant documents,'' in \emph{SIGIR '00: Proceedings of
  the 23rd annual international ACM SIGIR conference on Research and
  development in information retrieval}.\hskip 1em plus 0.5em minus 0.4em\relax
  New York, NY, USA: ACM, 2000, pp. 41--48.

\bibitem{cortes1995}
C.~{Cortes} and V.~{Vapnik}, ``Support-vector networks,'' \emph{Machine
  Learning}, vol.~20, no.~3, pp. 273--297, September 1995.

\bibitem{joachims09}
T.~Joachims, T.~Finley, and C.-N.~J. Yu, ``Cutting-plane training of structural
  svms,'' \emph{Mach. Learn.}, vol.~77, no.~1, pp. 27--59, 2009.

\bibitem{fei2005bayesian}
L.~Fei-Fei and P.~Perona, ``{A bayesian hierarchical model for learning natural
  scene categories},'' in \emph{IEEE Computer Society Conference on Computer
  Vision and Pattern Recognition (CVPR 2005)}, vol.~2, 2005.

\bibitem{sundaram2008}
S.~Sundaram and S.~Narayanan, ``Audio retrieval by latent perceptual
  indexing,'' in \emph{Acoustics, Speech and Signal Processing, 2008. ICASSP
  2008. IEEE International Conference on}, 2008, pp. 49 --52.

\bibitem{seyerlehner2008}
K.~Seyerlehner, G.~Widmer, and P.~Knees, ``Frame level audio similarity - a
  codebook approach,'' in \emph{Proceedings of the 11th International
  Conference on Digital Audio Effects (DAFx)}, Espoo, FI, 2008.

\bibitem{hoffman09}
M.~Hoffman, D.~Blei, and P.~Cook, ``Easy as {CBA}: A simple probabilistic model
  for tagging music,'' in \emph{Proceedings of the 10th International
  Conference on Music Information Retrieval}, 2009.

\bibitem{rabiner1993}
L.~Rabiner and B.-H. Juang, \emph{Fundamentals of speech recognition}.\hskip
  1em plus 0.5em minus 0.4em\relax Upper Saddle River, NJ, USA: Prentice-Hall,
  Inc., 1993.

\bibitem{buchanan2005}
C.~Buchanan, ``Semantic-based audio recognition and retrieval,'' Master's
  thesis, School of Informatics, University of Edinburgh, Edinburgh, U.K.,
  1995.

\bibitem{hartigan1975}
J.~A. Hartigan, \emph{Clustering Algorithms}, 99th~ed.\hskip 1em plus 0.5em
  minus 0.4em\relax New York, NY, USA: John Wiley \& Sons, Inc., 1975.

\bibitem{jebara2004}
\BIBentryALTinterwordspacing
T.~Jebara, R.~Kondor, and A.~Howard, ``Probability product kernels,'' \emph{J.
  Mach. Learn. Res.}, vol.~5, pp. 819--844, December 2004. [Online]. Available:
  \url{http://portal.acm.org/citation.cfm?id=1005332.1016786}
\BIBentrySTDinterwordspacing

\bibitem{bhattacharyya1943}
A.~Bhattacharyya, ``{On a measure of divergence between two statistical
  populations defined by probability distributions},'' \emph{Bull. Calcutta
  Math. Soc}, vol.~35, pp. 99--109, 1943.

\bibitem{scholkopf2002}
B.~Sch\"{o}lkopf and A.~Smola, \emph{Learning with Kernels}.\hskip 1em plus
  0.5em minus 0.4em\relax MIT Press, 2002.

\bibitem{barla2003}
A.~Barla, F.~Odone, and A.~Verri, ``Histogram intersection kernel for image
  classification,'' in \emph{Image Processing, 2003. ICIP 2003. Proceedings.
  2003 International Conference on}, vol.~3, 2003, pp. III -- 513--16 vol.2.

\bibitem{galleguillos2011}
C.~Galleguillos, B.~McFee, S.~Belongie, and G.~Lanckriet, ``{From region
  similarity to category discovery},'' in \emph{To appear in IEEE Conference on
  Computer Vision and Pattern Recognition (CVPR)}, 2011.

\bibitem{tingle10}
D.~Tingle, Y.~Kim, and D.~Turnbull, ``Exploring automatic music annotation with
  ``acoustically-objective'' tags,'' in \emph{IEEE International Conference on
  Multimedia Information Retrieval (MIR)}, 2010.

\bibitem{whitman2001}
B.~Whitman, G.~Flake, and S.~Lawrence, ``Artist detection in music with
  minnowmatch,'' in \emph{Neural Networks for Signal Processing XI, 2001.
  Proceedings of the 2001 IEEE Signal Processing Society Workshop}, 2001, pp.
  559--568.

\bibitem{dempster1977}
A.~P. Dempster, N.~M. Laird, and D.~B. Rubin, ``{Maximum Likelihood from
  Incomplete Data via the EM Algorithm},'' \emph{Journal of the Royal
  Statistical Society. Series B (Methodological)}, vol.~39, no.~1, pp. 1--38,
  1977.

\bibitem{jensen2007}
J.~Jensen, D.~Ellis, M.~Christensen, and S.~Jensen, ``Evaluation of distance
  measures between gaussian mixture models of {MFCC}s,'' in \emph{International
  Conference on Music Information Retrieval (ISMIR2007)}, 2007, pp. 107--108.

\bibitem{kullback1968}
S.~{Kullback}, \emph{{Information theory and statistics}}, {Kullback, S.}, Ed.,
  1968.

\bibitem{hershey2007}
J.~Hershey and P.~Olsen, ``Approximating the kullback leibler divergence
  between gaussian mixture models,'' in \emph{Acoustics, Speech and Signal
  Processing, 2007. ICASSP 2007. IEEE International Conference on}, vol.~4,
  2007, pp. IV--317 --IV--320.

\bibitem{tfidf}
G.~Salton and C.~Buckley, ``Term weighting approaches in automatic text
  retrieval,'' Ithaca, NY, USA, Tech. Rep., 1987.

\bibitem{sivic2003}
\BIBentryALTinterwordspacing
J.~Sivic and A.~Zisserman, ``Video google: A text retrieval approach to object
  matching in videos,'' in \emph{Proceedings of the Ninth IEEE International
  Conference on Computer Vision - Volume 2}, ser. ICCV '03.\hskip 1em plus
  0.5em minus 0.4em\relax Washington, DC, USA: IEEE Computer Society, 2003, pp.
  1470--. [Online]. Available:
  \url{http://portal.acm.org/citation.cfm?id=946247.946751}
\BIBentrySTDinterwordspacing

\bibitem{turnbull2008}
D.~Turnbull, L.~Barrington, D.~Torres, and G.~Lanckriet, ``Semantic annotation
  and retrieval of music and sound effects,'' \emph{Audio, Speech, and Language
  Processing, IEEE Transactions on}, vol.~16, no.~2, pp. 467 --476, feb. 2008.

\bibitem{turnbull2007}
\BIBentryALTinterwordspacing
------, ``Towards musical query-by-semantic-description using the cal500 data
  set,'' in \emph{Proceedings of the 30th annual international ACM SIGIR
  conference on Research and development in information retrieval}, ser. SIGIR
  '07.\hskip 1em plus 0.5em minus 0.4em\relax New York, NY, USA: ACM, 2007, pp.
  439--446. [Online]. Available:
  \url{http://doi.acm.org/10.1145/1277741.1277817}
\BIBentrySTDinterwordspacing

\bibitem{tsne}
L.~{van der Maaten} and G.~Hinton, ``Visualizing high-dimensional data using
  {t-SNE},'' \emph{Journal of Machine Learning Research}, vol.~9, pp.
  2579--2605, 2008.

\bibitem{rasiwasia2007}
N.~Rasiwasia, P.~Moreno, and N.~Vasconcelos, ``Bridging the gap: Query by
  semantic example,'' \emph{Multimedia, IEEE Transactions on}, vol.~9, no.~5,
  pp. 923 --938, aug. 2007.

\bibitem{barrington2007}
L.~Barrington, A.~Chan, D.~Turnbull, and G.~Lanckriet, ``Audio information
  retrieval using semantic similarity,'' in \emph{Acoustics, Speech and Signal
  Processing, 2007. ICASSP 2007. IEEE International Conference on}, vol.~2,
  april 2007, pp. II--725 --II--728.

\end{thebibliography}
\end{document}